\documentclass[showpacs,aps,twocolumn,nofootinbib]{revtex4-1}
\usepackage{epsfig}
\usepackage{graphicx}
\usepackage{amsmath,amssymb,amsfonts}
\usepackage{array}
\usepackage{url}
\usepackage{multirow}
\usepackage{natbib}
\usepackage{float}
\usepackage{lineno}
\usepackage{xspace}
\usepackage{ulem}
\usepackage{footnote}
\usepackage[usenames,dvipsnames]{color}
\definecolor{darkblue}{RGB}{0,0,196}
\usepackage[colorlinks=true,linkcolor=darkblue,citecolor=darkblue,urlcolor=darkblue]{hyperref}

\begin{document}
\title{Applicability of Hydrodynamics in Hadronic Phase of Heavy-Ion Collisions}
 
\author{Ronald Scaria}
\email[]{ronaldscaria.rony@gmail.com}
\author{Captain R. Singh}
\email[]{captainriturajsingh@gmail.com}
\author{Raghunath Sahoo}
\email[Corresponding author: ]{Raghunath.Sahoo@cern.ch}
\affiliation{Department of Physics, Indian Institute of Technology Indore, Simrol, Indore 453552, India}

\begin{abstract}
The hadronic phase and its dynamics in relativistic heavy-ion collisions are topics of immense discussion. The 
hadronic phase contains various massive hadrons with an abundance of the lightest hadron, i.e., $\pi$-mesons (pions).  In this paper,
 we consider that pions are in the thermal equilibrium in the hadronic phase and use second-order viscous hydrodynamics for a medium of massive pions to obtain its expansion to the boundary of the kinetic freeze-out. We achieve the kinetic freeze-out boundary with the Knudsen number $Kn>1$ limit. When this condition is met, hydrodynamics expansion breaks down, and the mean free path becomes sufficiently large in comparison with the system size so that the particle yields are preserved. Further, we investigate the effect of the massive fluid on the resonance particle yields, including re-scattering and regeneration, along with the natural decay widths of the resonances. The resonances can play an essential role in determining the characteristics of the hadronic phase as they have sufficiently small lifetimes, which may be comparable to the hadronic phase lifetime. In the current study, we predict the hadronic phase lifetime, which is further used to determine the $K^*(892)^0/K$, $\phi(1020)/K$, and $\rho(770)^0/\pi$ yield ratios at the kinetic freeze-out. We calculate these ratios as a function of charged particle multiplicity and transverse momentum and compare the findings with experimental data. Our calculations qualitatively agree with the experimental data, indicating a possible hydrodynamical evolution of the hadronic phase.

\end{abstract}
\date{\today}
\maketitle

\section{Introduction}
\label{Sec1}

Relativistic heavy-ion collisions provide a unique environment that encapsulates various phases of the QCD matter. The formation 
of a hot and dense, locally thermalized deconfined phase of colorful partons known as quark-gluon plasma (QGP) is expected 
to be formed in the very initial stages (proper time,  $\tau < 1 $ fm) following the collision. The emergence of a color-neutral hadronic 
phase ensues when the strongly interacting QGP phase dissipates, indicating a phase transition from a 
deconfined partonic phase to a colorless hadronic medium. Experimental data propose that the QGP phase exhibits behavior 
akin to those of a ``perfect fluid'' (PF) \cite{Arsene2005, Back2005, Adams2005, Adcox2005}, characterized by exceptionally 
 low shear viscosity ($\eta$) to entropy density ($s$) ratio near the AdS/CFT bound of $1/4\pi$~\cite{Kovtun2005}. The QGP,  formed 
 at an extremely high initial temperature, is presumed to evolve based on hydrodynamic models before cooling down to the chemical 
freeze-out temperature. At this temperature, all inelastic collisions cease, leading to the breakdown of QGP hydrodynamics. 
Many of these computations are carried out assuming the QGP phase to be both chemically and thermally equilibrated and 
composed of massless particles. The chemical freeze-out temperature closely aligns with the critical temperature ($T_c$) for 
phase transition predicted by lattice QCD (LQCD) calculations~\cite{Andronic:2017pug}. Using this framework, the lifetime of the QGP can be estimated in ultra-relativistic collisions~\cite{crs1,crs2}. However, defining the lifetime of the hadronic 
phase (the duration between chemical and kinetic freeze-outs) becomes somewhat ambiguous without an explicit definition of the kinetic freeze-out temperature or boundary.

Multiple models have been used to study the hadronic phase produced in relativistic nuclear collisions. 
 One of the earliest studies describes~\cite{bebie} the hadronic phase as a system that is thermally equilibrated but chemically decoupled. This phase allows momentum transfer between the constituent particles through collisions. 
The calculations have been performed by assuming that the yields of resonances stay in equilibrium with those of the daughter particles formed from their decays. This assumption constitutes the concept of partial chemical equilibrium (PCE). 
It has been shown that the particle production in nuclear collisions can be assumed to be from a fireball following this partial chemical equilibrium approach~\cite{Becattini:1996gy}. This concept has been further used to determine the kinetic freeze-out temperature from the yields of hadrons and resonances~\cite{Motornenko:2019jha}. Bayesian analysis of heavy ion data, on the other hand, has suggested that the particlization of the QGP phase occurs at about $150\text{--}160$ MeV range~\cite{Bernhard:2019bmu, Parkkila:2021yha, Bernhard:2016tnd, Parkkila:2021tqq}, where both chemical and thermal equilibria are considered to break down. As this limit is obtained from relative particle yields, one can think of this as a limit on the chemical freeze-out. The hadronic phase thus produced is generally treated using transport models like the Ultra-relativistic 
Quantum Molecular Dynamics (UrQMD) model~\cite{Petersen:2008dd} or the AMPT (A Multi-Phase Transport) model~\cite{Lin:2004en}. These are generally used as afterburners to the hydrodynamic calculations, as in 
Refs.~\cite{Steinheimer:2017vju, Knospe2016}. However, it has been observed that in such models, the intrinsic medium 
properties like viscosity are not well described~\cite{Song:2010aq, Demir:2008tr}. For example, the hadronic evolution in UrQMD is shown to remember the prior QGP properties at hadronization~\cite{Song:2010aq} and can lead to different evolutions for different initial conditions chosen, although the final state properties are reproduced. This may be credited to the absence of an adequate definition of intrinsic transport properties in the UrQMD hadronic phase. Also, in
 Ref.~\cite{Romatschke:2015dha}, it has been shown that the hadronic evolution itself has transport properties comparable 
with 
a low viscous fluid. Recent advancements in far-from-equilibrium hydrodynamics~\cite{Berges:2004ce, Romatschke:2017vte} 
also imply the possibility of rapid thermalization even if the system is away from equilibrium. Therefore, employing hydrodynamics could thus be an alternate approach to describing the hadronic phase. Hydrodynamics treats the system as a continuous, locally thermalized fluid, which allows for the use of macroscopic properties like temperature, pressure, and flow velocity to describe the evolution of the medium. This approach is more effective at capturing the collective flow and the evolution of thermodynamic quantities, which are essential for understanding the bulk properties of the system. Additionally, hydrodynamic models can naturally incorporate concepts like PCE and provide a framework for determining the hadronic phase lifetime. By maintaining a thermalized description 
throughout the evolution, the hydrodynamic description of the hadronic phase may offer a consistent depiction of the hadronic phase.

In this direction, the Knudsen number ($Kn$) can be used to study the possibility of thermalization and applicability of 
hydrodynamics in a medium. Low values of $Kn$ ($Kn \ll 1$) indicate systems with low gradients of hydrodynamic 
quantities, 
and thus the system is close to equilibrium (high degree of thermalization)~\cite{Romatschke:2017ejr}. A similar conclusion 
is also obtained in Ref.~\cite{Gallmeister:2018mcn}, where the applicability of fluid dynamics is strictly valid for cases 
of $Kn \sim 1$. 
In our previous study~\cite{Scaria:2022yrz}, we explored the possibility of fluid dynamics in the hadronic medium by 
studying $Kn$. We observed a considerably small value for $Kn$ at high temperatures within the excluded volume hadron 
resonance gas (EVHRG) formalism. In the current paper, we extend that study by explicitly deriving the valid equations 
for dissipative second-order (SO) hydrodynamics for a massive system and exploring the possibility of hydrodynamics in the 
hadronic phase by estimating the hadronic phase lifetime.

Short-lived resonances decay with a wide range of lifetimes, which allows them to be suitable probes for characterizing the 
properties of hadronic matter produced in heavy-ion collisions (see \cite{ALICE:2019xyr} and references therein) and also to 
estimate the hadronic phase lifetime. The resonances produced at the chemical freeze-out boundary can decay before the 
kinetic freeze-out, and their decay products undergo hadronic re-scattering, leading to a decrease in the resonance 
yields reconstructed at the kinetic freeze-out boundary. At the same time, pseudoelastic interactions between hadrons can regenerate the resonances in the medium, which increases the yields. Various transport and statistical thermal model calculations show that re-scattering and regeneration processes significantly affect the final state resonance yields~\cite{Andronic:2017pug, Singha2015, Knospe2016, Andronic2009, 
Cleymans1999, Chatterjee2015, Cho:2015qca}. These effects make them perfect probes to study the hadronic phase lifetime, as 
explored in Refs.~\cite{ALICE:2019xyr, Sahu:2019tch}. In this paper, we follow an alternate approach and use $Kn$ to govern the cooling of a hadronic medium from $T_c$ using a viscous second-order (0+1D) hydrodynamic model to obtain the hadronic phase lifetime. We introduce a model for the evolution of resonance particle yields during the hadronic phase. 
 The model is used within the ambit of hydrodynamic evolution to obtain the multiplicity and transverse momentum ($p_T$) dependent final yields, which are then compared with experimental data.

The paper is organized as follows.
Section~\ref{Sec2} details the equations derived for hydrodynamic calculations and the considerations that have been used to estimate the hadronic phase lifetime. Further, in Section~\ref{Sec3}, the formalism to obtain resonance yields at freeze out within a kinetic approach is detailed. In Section~\ref{Sec4}, we present the results along with 
discussions. Section~\ref{sum} summarizes the study. 
\section{Expanding Hadronic Phase}
\label{Sec2}
LQCD calculations predict the existence of a deconfinement transition from hadronic to partonic constituents for energy densities greater than 1 GeV fm$^{-3}$~\cite{Karsch2002}. At zero baryochemical potential ($\mu_{B}$), this deconfinement transition is predicted again by LQCD to be of crossover in nature~\cite{Aoki2006} at $T_{c} \simeq 155$~MeV~\cite{Bazavov2014, Borsanyi2010, Borsanyi2014}. The hadronic phase, thus predicted to be produced at the end of the QGP phase as a result of confinement, is further considered to expand and cool down. A sophisticated interplay of hadronic interactions, such as elastic and inelastic scattering, resonance production, decay, and other dynamical phenomena characterize the expanding hadronic phase. To precisely interpret the experimental data and deduce the properties of the QGP and the 
nature of the QCD phase transition, a thorough understanding of this complex phase is essential. It can provide unique insights into the fundamental properties of matter under extreme conditions, emphasizing the necessity for modeling this dynamic phase. In this Section, we explore the initial hadronization at chemical freeze-out and employ hydrodynamics to study the evolution of the hadronic medium and its cooling rate under various scenarios.

\subsection{Statistical Hadronization}
\label{sec2a}

The hadronization of the QCD matter is crucial to exploring QCD thermodynamics, and this transformation is effectively 
captured through the use of statistical thermal models, such as the ideal HRG model. Within the HRG model, the partition function for the $i$th hadron in the grand canonical ensemble is given by ~\cite{Andronic2012};
\begin{equation}
\label{eq1}
\ln Z^{id}_i = \pm \frac{Vg_i}{2\pi^2} \int_{0}^{\infty} p^2 dp\ln\{1\pm \exp[-(E_i-\mu_i)/T]\}
\end{equation}
where
$g_i$, 
$E_i = \sqrt{p^2 + m_i^2}$ (with $p$, the momentum of a particle), 
and $\mu_i$ are the degeneracy, energy, and chemical potential of the $i$th hadron, 
respectively. $V$ and $T$ define the volume and the temperature of the system, respectively. 
$\mu_i$ is given by 
\begin{equation}
\label{eq6}
\mu_i = B_i\mu_B + S_i\mu_S +Q_i\mu_Q, 
\end{equation}
{where $B_i$, $S_i$, and $Q_i$ are the baryon number, strangeness, and electric charge of the $i$th
hadron. 
Here, $\mu_{B}$, $\mu_{S}$,} $\mu_{Q}$ are the baryon, strange, and charge chemical \mbox{potentials, respectively.}

Knowing the partition function, one can calculate the pressure $P_i$, energy density $\varepsilon_i$, number density $n_i$, and entropy density 
$s_i$ as follows:
\begin{equation}
\label{eq2}
P^{id}_i(T,\mu_i) = \pm \frac{Tg_i}{2\pi^2} \int_{0}^{\infty} p^2 dp\ln\{1\pm \exp[-(E_i-\mu_i)/T]\},
\end{equation}
\begin{equation}
\label{eq3}
\varepsilon^{id}_i(T,\mu_i) = \frac{g_i}{2\pi^2} \int_{0}^{\infty} \frac{E_i\ p^2 dp}{\exp[(E_i-\mu_i)/T]\pm1},
\end{equation}
\begin{equation}
\label{eq4}
n^{id}_i(T,\mu_i) = \frac{g_i}{2\pi^2} \int_{0}^{\infty} \frac{p^2 dp}{\exp[(E_i-\mu_i)/T]\pm1},
\end{equation}
\begin{align}
 s^{id}_i(T,\mu_i)=&\pm\frac{g_i}{2\pi^2} \int_{0}^{\infty} p^2 dp \Big[\ln\{1\pm  \exp[-(E_i-\mu_i)/T]\}\nonumber\\ 
&\pm \frac{(E_i-\mu_i)/T}{\exp[(E_i-\mu_i)/T]\pm 1}\Big]
 \label{eq5}
 \end{align}

The ideal HRG model can be further extended to the EVHRG model by the inclusion of van der Waals (vDW) 
repulsive interactions between hadrons by considering a fixed volume to be occupied by each hadron.
If $v = 16\pi r_h^3/3$ is the volume occupied by each hadron where $r_h$ is the hard-core radius, then the 
 thermodynamic quantities in Equations \eqref{eq2}--\eqref{eq5} are given by~\cite{Andronic2012, Vovchenko2015}

\begin{equation}
\label{eq7}
P^{ev}(T,\mu) = \kappa P^{id}(T,\mu),
\end{equation}
\begin{equation}
\label{eq8}
\varepsilon^{ev}(T,\mu) = \frac{\kappa\varepsilon^{id}(T,\mu)}{1+\kappa vn^{id}(T,\mu)},
\end{equation}
\begin{equation}
\label{eq9}
 n^{ev}(T,\mu) = \frac{\kappa n^{id}(T,\mu)}{1+\kappa vn^{id}(T,\mu)},
\end{equation}
\begin{equation}
\label{eq10}
 s^{ev}(T,\mu) = \frac{\kappa s^{id}(T,\mu)}{1+\kappa vn^{id}(T,\mu)}, 
\end{equation}
where $\mu$ represents the chemical potential and $\kappa$ is the excluded volume suppression factor given by
\begin{equation}
\label{eq11}
\kappa = \exp\bigg(\frac{vP^{ev}}{T}\bigg) . 
\end{equation}

These models have been used extensively to study particle production in nucleus-nucleus collisions. From these studies, it has been found that particle yields in the final state can be explained with considerable precision by these models at temperatures that are in agreement with the phase 
transition temperature predicted by LQCD~\cite{Andronic:2005yp, Stachel:2013zma, Andronic:2017pug}. In the following, we use the EVHRG model to determine 
the initial yields considering the chemical freeze-out temperature to be equal to $T_{c} = 156$ MeV as given by the Wuppertal--Budapest Collaboration~\cite{Borsanyi2010, Borsanyi2014}.

\subsection{Temperature evolution of the hadronic medium}
\label{sec2b}
The hadronic medium produced at chemical freeze-out in nucleus-nucleus collisions is at a higher temperature and cools down with time. The temperature cooling can be studied by solving the first-order (FO) and SO fluid dynamical equations~\cite{Muronga:2003ta,Kouno:1989ps,Muronga:2001zk} derived within the framework 
of causal M\"uller--Israel--Stewart theory of dissipative fluid dynamics~\cite{Israel1976,Stewart1977,Israel1979}. 
In the (0+1D) scaling solution, one obtains~\cite{Muronga:2003ta,Muronga:2001zk}:
\begin{equation}
\label{eq12}
\frac{\partial\varepsilon}{\partial\tau} = -\frac{\varepsilon+P}{\tau} + \frac{\phi}{\tau}
\end{equation}

The PF, FO, and SO theories are distinguished by $\phi$, which defines the effect of viscous terms, as follows: 
\begin{equation}
\label{eq13}
    \phi = 0 
\;\;\; \text{(PF)
,}
\end{equation}
\begin{equation}
\label{eq14}
    \phi = \frac{4\eta}{3\tau} 
 \;\;\; \text{(FO),} 
\end{equation}
\begin{align}
\label{eq15}
\frac{d\phi}{d\tau} = -\frac{\phi}{\tau_{\phi}} - \frac{\phi}{2}\Big[\frac{1}{\tau} + \frac{1}{\beta_2} T \frac{d}{d\tau} \Big(\frac{\beta_2}{T} \Big) 
\Big]\nonumber\\ +\frac{2}{3\beta_2\tau} 
 \;\;\; \text{(SO),}
\end{align}
where $\tau_{\phi}$ is the relaxation time due to shear viscous component and $\beta_2 =  {\tau_{\phi}}/{(2\eta)}$. In the SO theory, $\phi$ indicates the characteristics of the system by measuring the change in $\eta$ through Equation~\eqref{eq15}. The solution of 
Equations~\eqref{eq12} and~\eqref{eq15} can be obtained numerically for a viscous system. The initial conditions for $\phi$ are calculated from the EVHRG model. 
Taking the AdS/CFT lower bound for $\eta/s = 1/(4\pi)$~\cite{Kovtun2005}, the value of $\phi$ at the time 
$\tau_c$  ($\tau$ at $T=T_c$) is found as $\phi_c =  {s_c}/{(3\pi\tau_c)}$ with $s_c = s(\tau_c)$. 

In Ref.~\cite{Prakash:1993bt}, it has been shown that for a hadronic medium, the equilibration times of kaons and protons are larger than that of pions. Also to be noted is that pions, being the least massive of all hadrons, are abundantly produced at the 
chemical freeze-out, making pions the dominant component of the hadron gas phase. 
Thus, in the present paper, we approximate the dynamics or evolution of the hadronic 
phase to be determined by pions~\cite{Prakash:1993bt}. We consider two different cases in this regard: (a) massless 
pionic medium and (b) massive pionic medium. 

The equations of system evolution for the massless case are already available in the literature~\cite{Muronga:2003ta, 
Muronga:2001zk} and are given by
\begin{equation}
\label{eq16}
\frac{dT}{d\tau} = -\frac{T}{3\tau} + \frac{\phi}{12aT^3\tau} ,
\end{equation}

\begin{equation}
\label{eq17}
\frac{d\phi}{d\tau} = -\sigma b T^3 \phi - \frac{1}{2}
 \left(\frac{1}{\tau} - 5\frac{1}{T}\frac{dT}{d\tau}\right
)
+\frac{8aT^4}{9\tau}, 
\end{equation}
where $\sigma$ represents the interaction cross-section between pions and the constants are given by  
$a=\pi^2/30$ and $b=3\zeta(3)/\pi^2$. Here $\zeta(\cdot)$ is the Euler--Riemann zeta function.

The cooling rate equations for the massive medium are derived in Section \ref{app_A} just below.
\subsection{Cooling Law for Massive Pionic Medium}
\label{app_A}

We consider the case of a massive pion gas in the Boltzmann limit. It is a known result that the pressure is proportional to the particle number density times the temperature~\cite{Vovchenko2015}:
\begin{equation}
    \label{A1}
    P = a'nT,
\end{equation}
where $a'$ is the proportionality constant which is explicitly obtained using Equations~\eqref{eq2} and~\eqref{eq4}. Now, using $P = c_s^2\varepsilon $, one obtains
\begin{equation}
    \label{A2}
    \varepsilon =  \frac{a'nT}{c_s^2} ,
\end{equation}
{where the speed of sound squared,   
 $c_s^2 = {\partial P}/{\partial \varepsilon}$, 
is calculated at each instant along the evolution considering a pionic 
medium.} 

On using Equations~\eqref{A1} and~\eqref{A2}, Equation~\eqref{eq12} is modified accordingly:
 \begin{equation}
    \label{A4}
    \frac{dT}{d\tau}
\left(T\frac{\partial n}{\partial T} + n - \frac{nT}{c_s^2}\frac{\partial c_s^2}{\partial T}
\right)
= -\frac{nT}{\tau}(c_s^2 + 
1) + \frac{\phi c_s^2}{a'\tau}, 
\end{equation}
where $
{\partial n}/{\partial T}$ 
can be obtained using Equation~\eqref{eq4}:
\begin{equation}
    \label{A5}
    \frac{\partial n}{\partial T} = \frac{\varepsilon}{T^2} = \frac{a'n}{c_s^2T}
\end{equation}

Thus, Equation~\eqref{A4} modifies to 
\begin{equation}
    \label{A6}
    \frac{dT}{d\tau}\left(
T\frac{a'n}{c_s^2T} + n - \frac{nT}{c_s^2}\frac{\partial c_s^2}{\partial T}\right)
= -\frac{nT}{\tau}(c_s^2 + 1) + \frac{\phi c_s^2}{a'\tau}, 
\end{equation}
which gives the temperature evolution for a massive viscous medium,
\begin{equation}
    \label{A7}
    \frac{dT}{d\tau} = -\frac{c_s^2T}{\tau}
\left(\frac{c_s^2 + 1}{c_s^2 + a' - T\frac{\partial c_s^2}{\partial T}}
\right) + 
\frac{\phi c_s^4}{a'n\tau\left(c_s^2 + a' - T\frac{\partial c_s^2}{\partial T}\right)}.
\end{equation}

In the case of a massive PF ($\phi = 0$), this equation reduces to
\begin{equation}
    \label{A8}
    \frac{dT}{d\tau} = -\frac{c_s^2T}{\tau}
\left(\frac{c_s^2 + 1}{c_s^2 + a' - T\frac{\partial c_s^2}{\partial T}}
\right) .
\end{equation}

It has to be noted here that on replacing $a' = 1$, $c_s^2 = 1/3$, and $P = aT^4$, Equation~\eqref{A7} reduces to 
Equation~\eqref{eq16}, which is the temperature cooling rate at the massless limit.

The effect of viscosity in Equation~\eqref{A7} is determined by solving Equation~\eqref{eq15}, which can be solved in the Boltzmann limit by making use of the results obtained in 
Refs.~\cite{Israel1976, Israel1979}. $\beta_2$ is generally defined as
\begin{equation}
    \label{B2}
    \beta_2 = \frac{1}{2\bar{\eta}^2P}
\left(1+\frac{6\bar{\eta}}{\beta}
\right),
\end{equation}
where $\beta = m/T$ and   $\bar{\eta} = {(\varepsilon+P)}/{(nm)} 
= {(c_s^2+1)a'T}/{m}$ 
with $m$ being the rest mass of the particle, so that 

\begin{equation}
    \label{B3}
    \beta_2 = \frac{1}{2}
 \left[
 \left(\frac{(c_s^2+1)a'T}{m}
 \right)^2a'nT
 \right]^{-1}
 \left(1+\frac{6(c_s^2+1)a'T^2}{m^2}
 \right) ,
\end{equation}
which gives,
\begin{equation}
    \label{B4}
    \beta_2 = \frac{1}{2} \,
 \frac{m^2+6(c_s^2+1)a'T^2}{(c_s^2+1)^2(a'T)^3n}
 .
\end{equation}

From Equation~\eqref{B4}, 

\begin{equation}
    \label{B5}
    \frac{\beta_2}{T} = \frac{1}{2a'^2n(c_s^2+1)}
 \left(
\frac{m^2}{T^4(c_s^2+1)a'} + \frac{6}{T^2}\right)
.
\end{equation}

Using chain rule to solve Equation~\eqref{eq15}), one finds
\begin{equation}
    \label{B6}
    \frac{d}{d\tau}
 \left(\frac{\beta_2}{T}
 \right) = \frac{d}{dT}
 \left(\frac{\beta_2}{T}
 \right) \frac{dT}{d\tau}, 
\end{equation}
where ${dT}/{d\tau}$ is given by Equation \eqref{A6}. From Equation~\eqref{B5},
\begin{align}
\label{B7}
\frac{d}{dT}
 \left(\frac{\beta_2}{T}
 \right) = \frac{1}{2a'^2n(c_s^2+1)}
 \left(-\frac{4m^2}{T^5(c_s^2+1)a'}-\frac{12}{T^3}\right) 
\nonumber\\ 
+ 
 \left(\frac{m^2}{T^4(c_s^2+1)a'}+\frac{6}{T^2}\right)
 \times
 \left(
-\frac{1}{2a'^2n^2(c_s^2+1)}\frac{\partial n}{\partial T}
 \right)
 .
\end{align}
 
Then, using
Equation~\eqref{A5}, one obtains
\begin{align}
\label{B8}
\frac{d}{dT}
 \left(\frac{\beta_2}{T}
 \right) 
= \frac{1}{2a'^2n(c_s^2+1)}
 \left(
-\frac{4m^2}{T^5(c_s^2+1)a'}-\frac{12}{T^3}
\right)
\nonumber\\ +\frac{1}{2a'nc_s^2(c_s^2+1)}
 \left(
-\frac{m^2}{T^5(c_s^2+1)a'}-\frac{6}{T^3}
 \right) ,
\end{align}
which further reduces to
\begin{align}
\label{B9}
\frac{d}{dT}
 \left(\frac{\beta_2}{T}
 \right)
= -\frac{1}{2a'n(c_s^2+1)T^3}\left[ 
\frac{12}{a'}+\frac{6}{c_s^2}
 \right.\nonumber\\ + 
 \left. 
  \frac{m^2}{T^2(c_s^2+1)a'}
 \left(\frac{4}{a'}+\frac{1}{c_s^2}
 \right)
 \right]
\end{align}

Using Equations~\eqref{B4},~\eqref{B6}, and~\eqref{B9} in Equation~\eqref{eq15} gives
\begin{widetext}
\begin{equation}
\label{B10}
\frac{d\phi}{d\tau} = -\frac{\phi}{\tau_{\phi}} - \frac{\phi}{2}\Big[\frac{1}{\tau} - \frac{T(c_s^2+1)a'^2}{m^2+6(c_s^2+1)a'T^2}\Big\{\frac{12}{a'}+\frac{6}{c_s^2} + \frac{m^2}{T^2(c_s^2+1)a'}\Big(\frac{4}{a'}+\frac{1}{c_s^2}\Big)\Big\}\frac{dT}{d\tau} \Big] +\frac{4}{3\tau}\frac{T^3(c_s^2+1)^2a'^3n}{m^2+6(c_s^2+1)a'T^2}
\end{equation}
\end{widetext}

\begin{figure}[H]
\includegraphics[scale = 0.42]{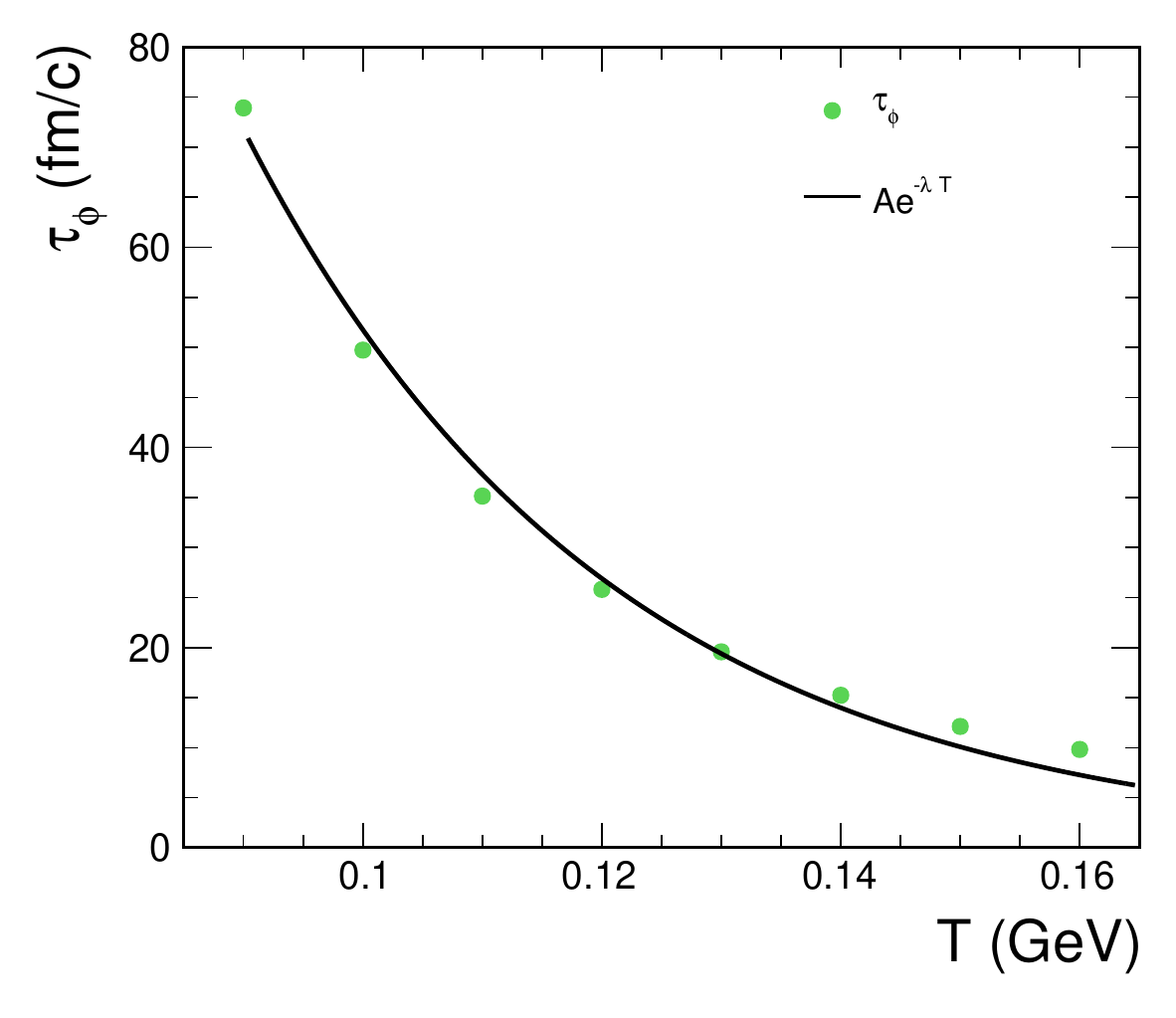}
\caption{Relaxation time $\tau_\phi$ of a massive pion gas as a function of temperature $T$. 
 The fit parameters obtained are $A = 1362.73\pm 212.198$ fm/c and $\lambda = 
32.7087\pm 1.5375$ GeV$^{-1}$.}
\label{figphi}
\end{figure}

The only unknown quantity in Equation~\eqref{B10} is $\tau_\phi$.  The particle-dependent relaxation time for a hadron gas can 
be obtained using the Boltzmann transport equation (BTE)~\cite{Kadam:2015xsa, Tiwari2018, Scaria:2022yrz}. For a scattering of $i$th and $j$th particles, $i(p_i) + j(p_j) \xrightarrow{} i(p_k) + j(p_l) $, the energy-dependent relaxation time
 is given by 
\begin{equation}
    \label{B11a}
    {\tau_\phi}^{-1}(E_i) = \sum_{jkl}\int \frac{d^3p_jd^3p_kd^3p_l}{512\pi^9}W(i,j \xrightarrow{} k,l) f_j^0 , 
\end{equation}
where $W(i,j \xrightarrow{} k,l)$ is the transition rate and $f_j^0$ is the distribution function. In the 
center-of-mass frame, Equation \eqref{B11a} reduces to   
\begin{equation}
    \label{B11b}
    {\tau_\phi}^{-1}(E_i) = \sum_{j}\int \frac{d^3p_j}{8\pi^3} \sigma_{ij}v_{ij} f_j^0 , 
\end{equation}
where $\sigma_{ij}$ is the interaction cross section and $v_{ij}$ is the relative velocity. Equation~\eqref{B11b} can be 
averaged over $f_i^0$ to obtain the averaged relaxation time $\Tilde{\tau}_{i\phi}$ for particle $i$:
\begin{align}
    \label{B11}
    \Tilde{\tau}_{i\phi}^{-1} &= \frac{\int \frac{d^3p_i}{8\pi^3} \tau^{-1}(E_i) f_i^0}{\int \frac{d^3p_i}{8\pi^3} f_i^0} \nonumber\\
    &= \sum_j\frac{\int \frac{d^3p_i}{8\pi^3} \frac{d^3p_j}{8\pi^3} \sigma_{ij}v_{ij} f_i^0 f_j^0}{\int \frac{d^3p_i}{8\pi^3} f_i^0}  \nonumber\\
    &= \sum_jn_j\langle\sigma_{ij}v_{ij}\rangle .
\end{align}

The thermal averaged cross-section times relative velocity is given by~\cite{Kadam:2015xsa, Tiwari2018, Scaria:2022yrz};

\begin{widetext}
\begin{equation}
    \label{eq27}
    \langle\sigma_{ij}v_{ij}\rangle = \frac{\sigma\int p_ip_jE_i~dE_i~E_j~dE_j~d\cos\theta ~f^0_if^0_j\times \frac{\sqrt{(E_i~E_j-p_ip_j\cos \theta)^2-
    (m_im_j)^2}}{E_i~E_j-p_ip_j\cos\theta}}{\int p_ip_jE_i~dE_i~E_j~dE_j~d\cos\theta~ f^0_if^0_j}
\end{equation}
\end{widetext}
where $E_i$ and $E_j$ are particles' energies and $\theta$ is the angle between the particles' momenta $p_i$ and $p_j$. 
The integration over $\cos\theta$  is taken from 
$-$1 to +1.  
$f_i^0$ in Equation \eqref{eq27} is given by 
\begin{equation} 
    \label{eq28}
    f_i^0 = \exp\left(-\frac{E_{i}-\mu_B
}{T}\right) ,
\end{equation}
where $\mu_B$ is set to zero in our calculations. 

Equation~\eqref{B11} is used to find the pion relaxation times at different temperatures and the values obtained are 
fitted by an exponent 
$\tau_\phi = A\exp(-\lambda T)$
, as shown in Figure~\ref{figphi}, to obtain an approximate dependence on temperature. This fit function 
is further used in Equation~\eqref{B10}.

Equations~\eqref{A8} and~\eqref{B10} describe the cooling rate for massive pion gas. In our calculations, we assume $m_{\pi} = 0.139$ GeV.


\subsection{Transverse Correction to the Cooling Law}
The effect of transverse expansion can be included in the (0+1D) scaling solution for the partonic state by considering that cooling 
is defined by the transverse expanded time $\tau_{tr}>\tau$, {i.e.,} $T_{tr}(\tau) < T(\tau)$ ~\cite{crs1,crs2, Lokhtin:1996ht}. 
 This also implies that the temperature cooling begins at a lower temperature $T_{tr}(\tau_{0}) < 
T_{0}(\tau_0)$, with $\tau_0$ being the initial thermalization time. Such a consideration makes cooling faster due to the effect of transverse expansion. Thus, due to the transverse expansion from the initial phase of collision, the system achieves chemical 
freeze-out temperature, $T_{c}$, earlier than in a simplified  (0+1D) expansion.

Let us define chemical freeze-out time as $\tau_{QGP}$. In the case of ideal expansion without transverse cooling, one may roughly assume that the QGP fireball lifetime $\tau_{QGP}^{id} = R_T$ ~\cite{Hanus:2019fnc}, where $R_T$ is the transverse radius obtained from the MC Glauber model~\cite{Loizides:2017ack}. Taking a similar approach but with the transverse expansion, one may assume the hadronization to take place 
when $\tau_{tr} = R_T$. Now, defining $\tau_{tr}\cong \tau_{QGP}^{tr} + \frac{R_{T}}{c_s}\frac{\sqrt{2}-1}{\sqrt{2}}$, 
$\tau_{QGP}^{tr}$ can be straightforwardly determined~\cite{Lokhtin:1996ht}. $\tau_{QGP}^{tr}$ is further used as the hadronization time $\tau_c$ defined earlier as the initial time $\tau_0$ for the hydrodynamical evolution of the hadronic phase.

\begin{figure*}
\includegraphics[scale = 0.42]{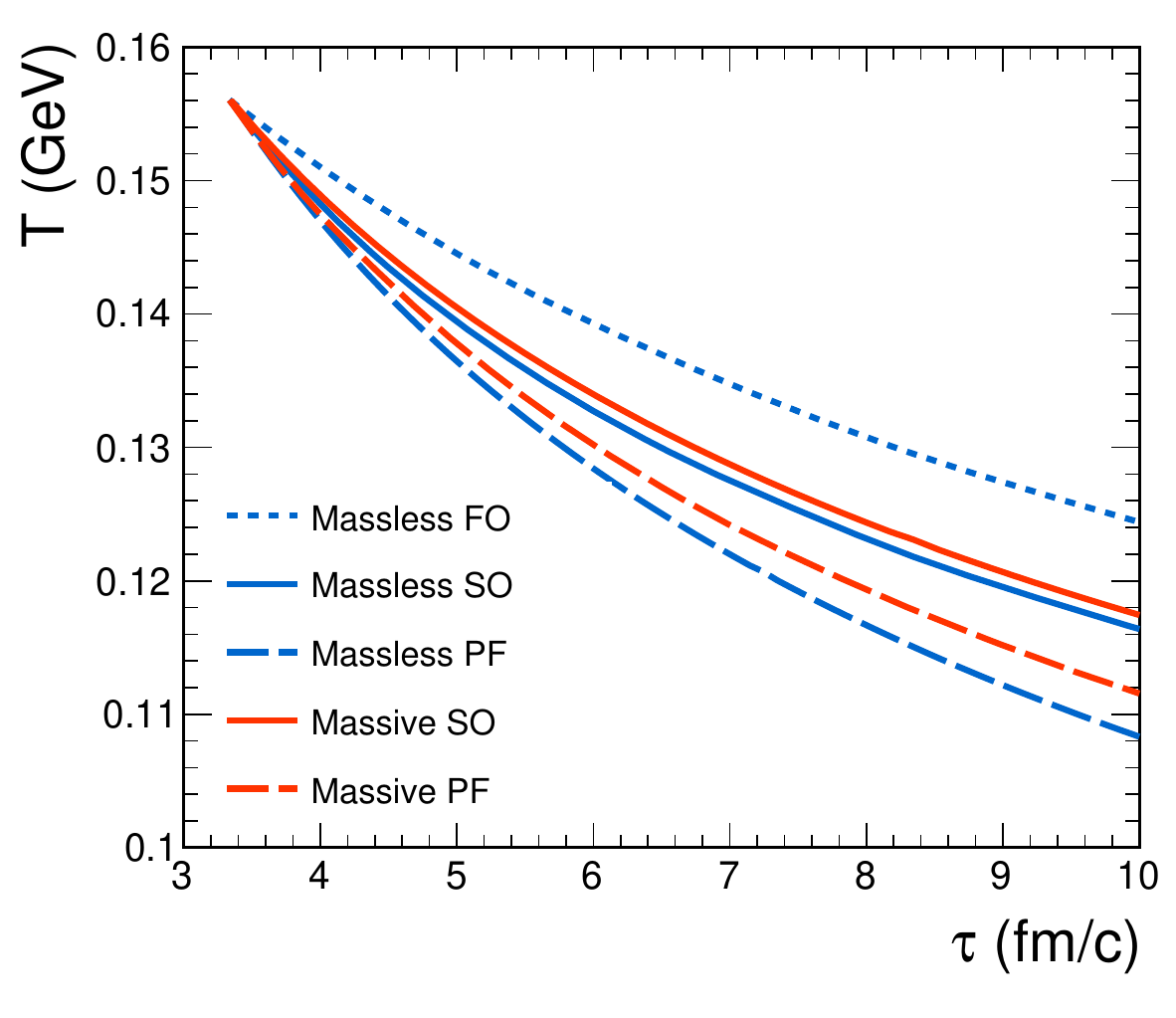}
\includegraphics[scale = 0.42]{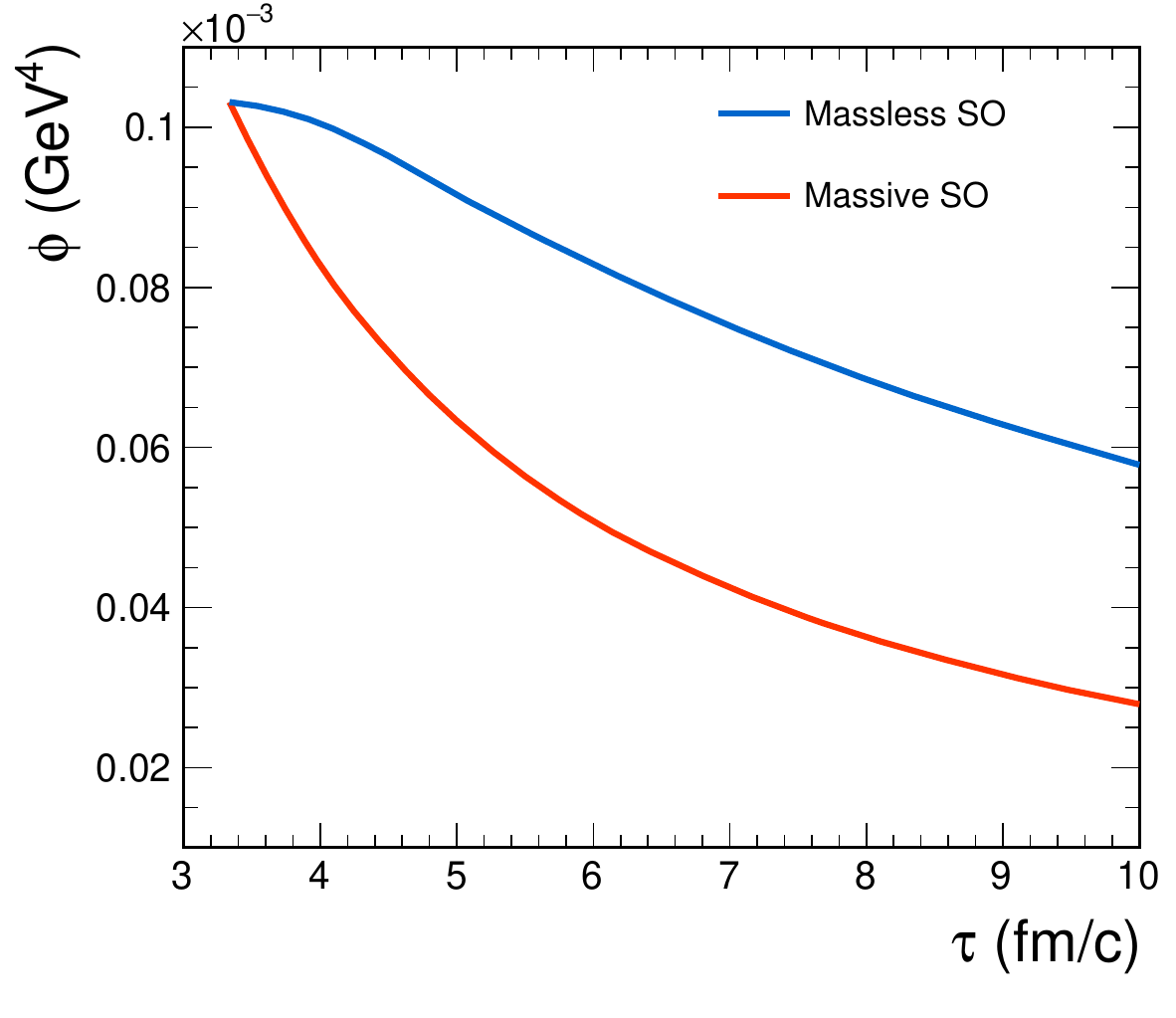}
\caption{\textbf{Left}: temperature evolution as a function of proper time for both massive and massless constituent pions 
as calculated for the first-order (FO), and second-order (SO) hydrodynamical theories and the perfect liquid (PF) limit. 
The initial temperature is taken as $T_c   = 0.156$ GeV to start the cooling law from the chemical freeze-out~\cite{Borsanyi2010, Borsanyi2014}.
 \textbf{Right}: comparison of the viscous component as a function of proper time for massive and massless SO  
hydrodynamics. See text for details. }
\label{fig1}
\end{figure*}

Figure~\ref{fig1} (left) compares the temperature evolution of the massive and massless cases of PF and SO hydrodynamic models along with the FO massless case. One can see that viscous terms make cooling slower. The SO theory leads to 
a much faster cooling rate than the FO theory. The effect of particle mass makes cooling slower, as observed in the PF 
limit. Figure~\ref{fig1} (right) shows the impact of particle mass on the viscous component, which slows down cooling by generating more heat. 
In the case of massless particles, viscosity is higher and decays slowly as compared to the massive case. 
Thus, slowing down of the cooling for massive cases is opposed by the lower viscosity in the SO
hydrodynamics as compared to the massless case. An interplay of these effects can be seen in the massive 
SO theory results in Figure~\ref{fig1} (left), which is close enough to the massless condition albeit slightly slower. 
In the following, we consider only the SO model cooling scenario, because the FO theory solution is acausal and leads to 
unstable results~\cite{Muronga:2003ta}.

The validity of hydrodynamics is governed by $Kn$.  $Kn$ describes the probability of collisions between the particles in 
the system. $Kn \ll 1$ indicates that the system size is larger than the mean free path, 
indicating thermalization. Therefore, it becomes feasible to study the medium evolution under hydrodynamics. 

$Kn$ is defined as
\begin{equation}
\label{eq20}
{Kn} = L/D, 
\end{equation}
where $L  =  {1}/{(\sqrt{2}n\sigma)}$   
is the mean free path of the system, $n$ and $\sigma=4\pi r_h^2$ are the number density and the interaction 
cross-section between the particles of the medium, respectively.  $D = 2R$ is the characteristic system size,  where
 $R$ denotes the system transverse radius. Hydrodynamic evolution ends and the
equilibrium is considered to break down at the moment when the mean free path becomes greater than the system size, 
{i.e.,} at $Kn > 1$~\cite{Gallmeister:2018mcn}. The time-dependent transverse radius used in Equation~\eqref{eq20} can be obtained following the method 
given in Ref.~\cite{Lokhtin:1996ht}, where the effective radius at a 
time $\tau > \tau_i$ is given by 
\begin{equation}
\label{eq19}
R\cong R_i
 \left(
 1+\frac{1}{6}\sqrt{\frac{1}{\tau_i}}\frac{(\tau-\tau_i)^{3/2}}{R_i}
 \right), 
\end{equation}
where $\tau_i$ is the proper time at the previous instant.

\section{Kinetic Formation Model for Resonance Particles}
\label{Sec3}
At the end of the QGP phase, the deconfined quarks and gluons are combined to form hadrons, entering the hadronic 
phase. As a result of confinement, physical observables such as particle spectra, flow, fluctuations and correlations, etc., in QCD are required to be defined in terms of hadrons. The dynamics of 
interactions within the hadronic phase can affect these observables. 
These interactions involve both elastic scatterings, which change the momentum distribution without altering the particle composition, and inelastic processes, such as resonance formation and decay, that modify the chemical composition of the hadronic matter. This transport and evolution of hadrons in the hadronic phase strongly influence the final particle yields. 

Several resonance particles are produced in the final state of proton--proton and heavy-ion collisions with lifetimes varying from 1.1 fm/$c$ (for $f_2(1270)$) to 46.3 fm/$c$ (for $\phi(1020)$), with $c$ denoting the speed of light. 
 Such a wide range in the lifetimes provides an exceptional opportunity to use them as probes to study the system formed in 
relativistic nuclear collisions~\cite{Bleicher:2002dm, Torrieri:2001ue, Johnson:1999fv, Ilner:2017tab, ALICE:2019xyr}. 
The dynamics of these resonances within the hadronic phase significantly affect the yields and momentum distributions of the stable final state particles like pions, kaons and protons. Long-lived particles like $\eta$ (with the lifetime $\tau_l\sim 10^{-19}$ s) and $\eta'$ ($\tau_l\sim 10^{-21}$ s) mesons can also affect the final state pion yield obtained at detectors through their decay channels.
In the subsequent calculations, we used three different resonance particles---namely, $\rho(770)^{0}$, $K^*(892)^0$, and $\phi(1020)$. Out of these, the $K^*(892)^0$ meson occupies a sweet spot, with a lifetime of 4.16~fm/$c$, making it a 
suitable probe to study the hadronic phase and hadronic phase lifetime, as performed in \cite{ALICE:2019xyr}.
If the resonance decays within the hadronic phase, then the decay products are subject to interactions before kinetic freeze out, which alters their momentum 
and prevents their reconstruction (re-scattering) ($\rho(770)^0\xrightarrow{}\pi^+\pi^-$, $K^*(892)^0\xrightarrow{}\pi K$, $\phi(1020)\xrightarrow{}K^+K^-$). Similarly, hadrons in the medium can interact together to form 
resonances that may live up to the kinetic freeze-out allowing a successful reconstruction (regeneration) ($\pi^+\pi^-\xrightarrow{}\rho(770)^0$, $\pi K\xrightarrow{}K^*(892)^0$, $K^+K^-\xrightarrow{}\phi(1020)$). 
To obtain an experimental estimate of the hadronic phase lifetime, a natural decay formalism of $K^*(890)^0$ mesons is considered as given by~\cite{ALICE:2019xyr,Sahu:2019tch}
\begin{equation}
 \left[\frac{K^*}{K}
 \right]_{\rm kin} = 
 \left[\frac{K^*}{K}
 \right]_{\rm chem} \exp
 \left(\frac{\tau_H}{\tau_l}
 \right) 
,
    \label{expkstar}
\end{equation}
where $\tau_H$ is the hadronic phase lifetime and the subscripts ``kin''  and ``chem'' denote the kinetic and chemical freeze-out boundaries, respectively.  This formalism considers only rescattering due to decay and does not take into consideration the regeneration effects described above.
Thus, it is imperative to include both of these effects in the calculations while studying the final resonance yields.

In this respect, we use a modified version of the kinetic formation model used to explain the yield of $J/\psi$ meson in the final 
state~\cite{Thews:2001hy, Thews:2000rj}. This model was subsequently modified to include transverse momentum dependence of the decay and regeneration cross-sections~\cite{crs1,crs2}. According to this model, the final state yield of resonance particles ($A\rightleftharpoons BC$) is given by
\begin{align}
N_f^A(\tau_f,p_T) = \nonumber\\\epsilon(\tau_f,p_T)\lambda_D(\tau_f,p_T)\big[N_i^A(\tau_c,p_T) + N^B(p_T) N^C(p_T) 
\nonumber\\\times\int_{\tau_c}^{\tau_f}\Gamma_F(p_T)[V(\tau)\epsilon(\tau,p_T)\lambda_D(\tau,p_T)]^{-1}d\tau \big] ,
    \label{eq21}
\end{align}
where $\lambda_D(\tau,p_T)$ and $\epsilon(\tau,p_T)$ give the contribution due to natural decay and co-moving hadrons, respectively, to re-scattering at time $\tau$ while $\Gamma_F(p_T)$ gives the effect of regeneration. 
$\tau_f$ denotes the kinetic freeze-out time.  The quantities $N_i^A(\tau_c,p_T)$, $N^B(p_T)$ and $N^C(p_T)$ represent 
the $p_T$-dependent numbers of resonance particles and their decay products at chemical freeze-out determined through the EVHRG model. 
$N^B(p_T)$ and $N^C(p_T)$ are estimated such that their combination can produce a resonance particle at a given $p_T$.


The $V(\tau)$ in Equation~\eqref{eq21} denotes the time-dependent volume of the evolving system. It is obtained considering the isentropic evolution of the hadronic phase. Based on the yields of weakly bound nuclei in the final states, it was suggested that any further evolution of the fireball after $T_{c}$ has to be close to isentropic~\cite{Andronic:2010qu, Andronic:2017pug}. 
Further, in Ref.~\cite{Hanus:2019fnc}, it was quantitatively shown that for viscous hydrodynamic evolution, entropy production is minimal for the extended period when freeze-out is introduced. Thus, the volume profile $V(\tau,b)$ of the medium is obtained as follows:
\begin{equation}
\label{eq18}
V(\tau,b) = \frac{V_c(\tau_c,b)s_c(\tau_c)}{s(\tau)}, 
\end{equation}
where $V_c(\tau_c,b) = \tau_{tr}A_T(b)$ is the initial volume at $\tau_0$ with $b$ the impact parameter and 
$A_T$ being the transverse overlap area obtained from the MC Glauber model~\cite{Loizides:2017ack}.

\subsubsection{Regeneration}
Regeneration of resonance particles in the hadronic phase is modeled in Equation~\eqref{eq21} through $\Gamma_F(p_T)$, which is given by
\begin{equation}
    \label{eq25}
    \Gamma_F(p_T) = \langle\sigma_{reg}^A(p_T)v_{rel}\rangle_{B,C} \, , 
\end{equation}
where the averaging $\langle\cdots\rangle_{B,C}$ is taken over the $B$ and $C$ particles as is prescribed in Equation \eqref{eq27},  $\sigma_{reg}(p_T)$ is the regeneration cross-section, and $v_{rel}$ is the relative velocity between the particles. For 
$\rho(770)^0$ and $K^*(892)^0$ mesons, the temperature-dependent regeneration cross-sections ($\sigma_{reg}(T)$) are given 
by~\cite{Yang:2017nmv}
\begin{align}
    \sigma_{reg}^{\rho}(T) = 54.1\Big(\frac{T/T_c-0.42}{0.18}\Big)^{3.3} \exp\Big[ 3.3\Big(1-\frac{T/T_c-0.42}{0.18}\Big)\Big] \nonumber\\+5.32\Big(\frac{T/T_c-
    0.42}{0.5}\Big)^{84} \exp\Big[ 84\Big(1-\frac{T/T_c-0.42}{0.5}\Big)\Big]
    \label{eq26}
\end{align}
\begin{align}
    \sigma_{reg}^{K^*}(T) = 42.2\Big(\frac{T/T_c-0.5}{0.115}\Big)^{1.83} \exp\Big[ 1.83\Big(1-\frac{T/T_c-0.5}{0.115}\Big)\Big] \nonumber\\+4.95\Big(\frac{T/T_c-
    0.5}{0.407}\Big)^{50} \exp\Big[ 50\Big(1-\frac{T/T_c-0.5}{0.407}\Big)\Big]
    \label{eq26a}
\end{align}
where the critical temperature was taken to be $T_c = 0.175$ GeV. In the absence of such temperature-dependent parameterizations for $\phi(1020)$ mesons, we used the vacuum cross-section of 8.05 mb~\cite{Li:2020jdg} at all temperatures, 
i.e., $\sigma_{reg}^{\phi}(T) = 8.05$ mb.

The temperature experienced by the resonance particle can differ from the medium temperature. Thus, we use the relativistic Doppler shift due to the relative velocity 
 $v_r$ of resonance as compared to the medium to obtain the effective temperature felt by the resonance particle. The effective temperature is given 
by~\cite{crs1,Nendzig:2014qka,Hoelck:2016tqf}
\begin{equation}
    \label{eq43}
    T_{\rm eff}(\tau,p_T) = T(\tau) ~\frac{\sqrt{1-|v_r|^2}}{2|v_r|}~ \ln
 \left( \frac{1+|v_r|}{1-|v_r|}
 \right) .
\end{equation}

To calculate $v_r$, we used the medium velocity $v_m = 0.5 c$ comparable to the expansion velocity obtained from blastwave 
fits to data~\cite{ALICE:2019hno} and the 
resonance velocity $v_A = p_T/E_T$ with the transverse energy $E_T = \sqrt{p_T^2 + m_A^2}$, where $m_A$ is the mass of the particle $A$. Thus, the $p_T$-dependent regeneration cross-sections $\sigma_{reg}^A(p_T)$ are obtained from the 
temperature-dependent cross-sections $\sigma_{reg}^A(T)$ by using $T = T_{\rm eff}(\tau,p_T)$.

\subsubsection{Decay}
The re-scattering effect introduced in Equation~\eqref{eq21} was modeled to include two different effects. Resonance particles produced at chemical freeze-out can undergo natural decay ($A\xrightarrow{}BC$). They might also decay as a result of interaction with another particle ($X$) within the medium (co-moving hadron): $AX\xrightarrow{}BCX$. We consider that the resonance particle is re-scattered in both cases. The contribution due to 
natural decay is given by
\begin{equation}
    \label{eq22}
    \lambda_D(\tau,p_T) = \exp
 \left(-\frac{\tau-\tau_c}{\tau_l(p_T)}
 \right) 
,
\end{equation}
 where $\tau_l(p_T) = \tau_{reso}\gamma(p_T)$ is a product of the vacuum lifetime $\tau_{reso}$ of the resonance particle and the $p_T$-dependent Lorentz factor, the latter, in our calculations considered to read  $\gamma(p_T) = \sqrt{1 + (p_T/mc)^2}$ \cite{ALICE:2019xyr}. 
 The effect of co-moving hadrons is modeled based on the kinetic model~\cite{Thews:2001hy, Thews:2000rj}: 
\begin{equation}
    \label{eq23}
    \epsilon(\tau,p_T) = \exp
\left[-\int_{\tau_c}^{\tau}\Gamma_D(p_T) N_{\rm co} [V(\tau)]^{-1}d\tau
\right] 
 , 
\end{equation}
where $N_{\rm co}$ represents the total number of the co-moving hadrons and $\Gamma_D(p_T)$ is the decay rate of the resonances averaged over the co-moving hadron momentum distribution and is given by
\begin{equation}
    \label{eq24}
    \Gamma_D(p_T) = \langle\sigma_{\rm co}^A(p_T)v_{rel}\rangle_{A,{\rm co}} ,
\end{equation}
where $\sigma_{\rm co}^A(p_T)$ is the $p_T$-dependent decay cross-section. This decay cross-section is obtained by using detailed balance from the regeneration cross-section $\sigma_{reg}^A(p_T)$~\cite{Thews:2000rj,crs1}. 
$\langle \sigma v \rangle$ term in Equation \eqref{eq24} is obtained for particle $A$ and the co-moving hadron using Equation~\eqref{eq27}.

\section{Results and Discussion}
\label{Sec4}

The hadronic medium considered in this paper, consists of all hadrons and resonances with masses up to 2.25 GeV given in the Review of Particle Physics book~\cite{Zyla2020}. Since the properties of the medium depend on the parameter $r_h$, it has to be chosen appropriately. The repulsive 
interactions being mediated by $\omega$ meson and the range of interaction being inversely proportional to the mass of the mediator, we have chosen the 
hard-core radius, $r_h = 0.25$ fm to incorporate repulsive interactions. This radius is chosen to be uniform for all hadrons.

\begin{figure}[htp]
\includegraphics[scale = 0.42]{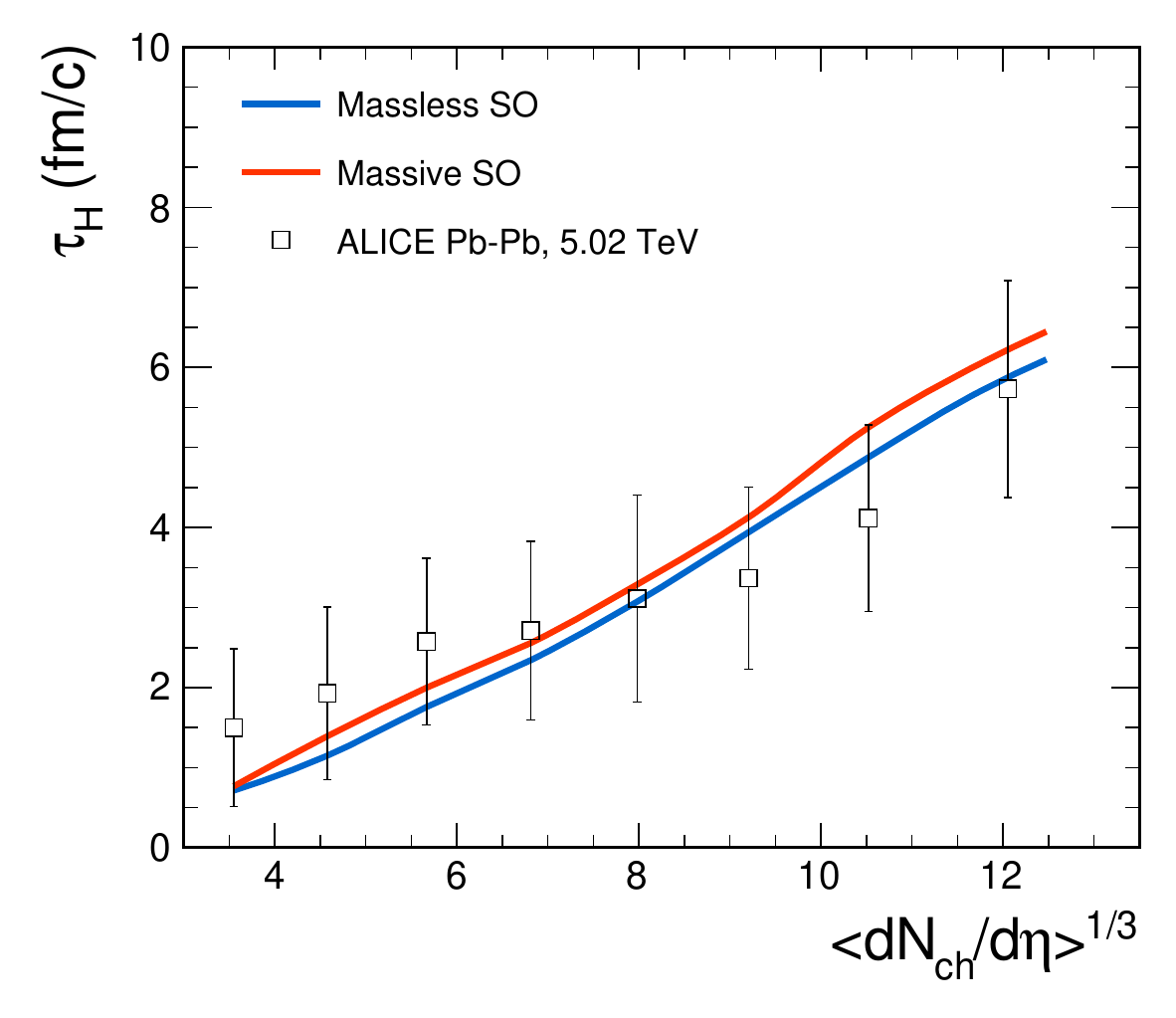}\vspace{-12pt}

\caption{Hadronic phase lifetime obtained from the calculations are compared with the results obtained from Pb--Pb collisions at the nucleon--nucleon center-of-mass energy $\sqrt{s_{\rm NN}}$ = 5.02 TeV by the ALICE~experiment  \cite{ALICE:2019xyr}.}
\label{fig2}
\end{figure}

The Knudsen number is an essential parameter as it indicates the applicability of hydrodynamics in the medium. As discussed in 
Section \ref{Sec2}, we used this parameter to obtain the endpoint $\tau_f$ for hydrodynamics.  
Finding $\tau_f$ allows us to estimate the hadronic phase lifetime $\tau_H$ just by calculating the difference between the 
initial ($\tau_c$) and final ($\tau_f$) time. In Figure~\ref{fig2}, we depict the hadronic phase lifetime obtained in our calculations 
for both the massive and massless SO hydrodynamics and compare it with the 5.02 TeV Pb--Pb collision results from the 
LHC~\cite{ALICE:2019xyr}. The massive SO theory predicting a slightly slower cooling rate than the massless case (as seen in Figure~\ref{fig1}), gives a higher lifetime. A reasonable agreement is observed between our calculations and 
the experimental results in both cases.

\begin{figure*}[htp]
\includegraphics[scale = 0.28]{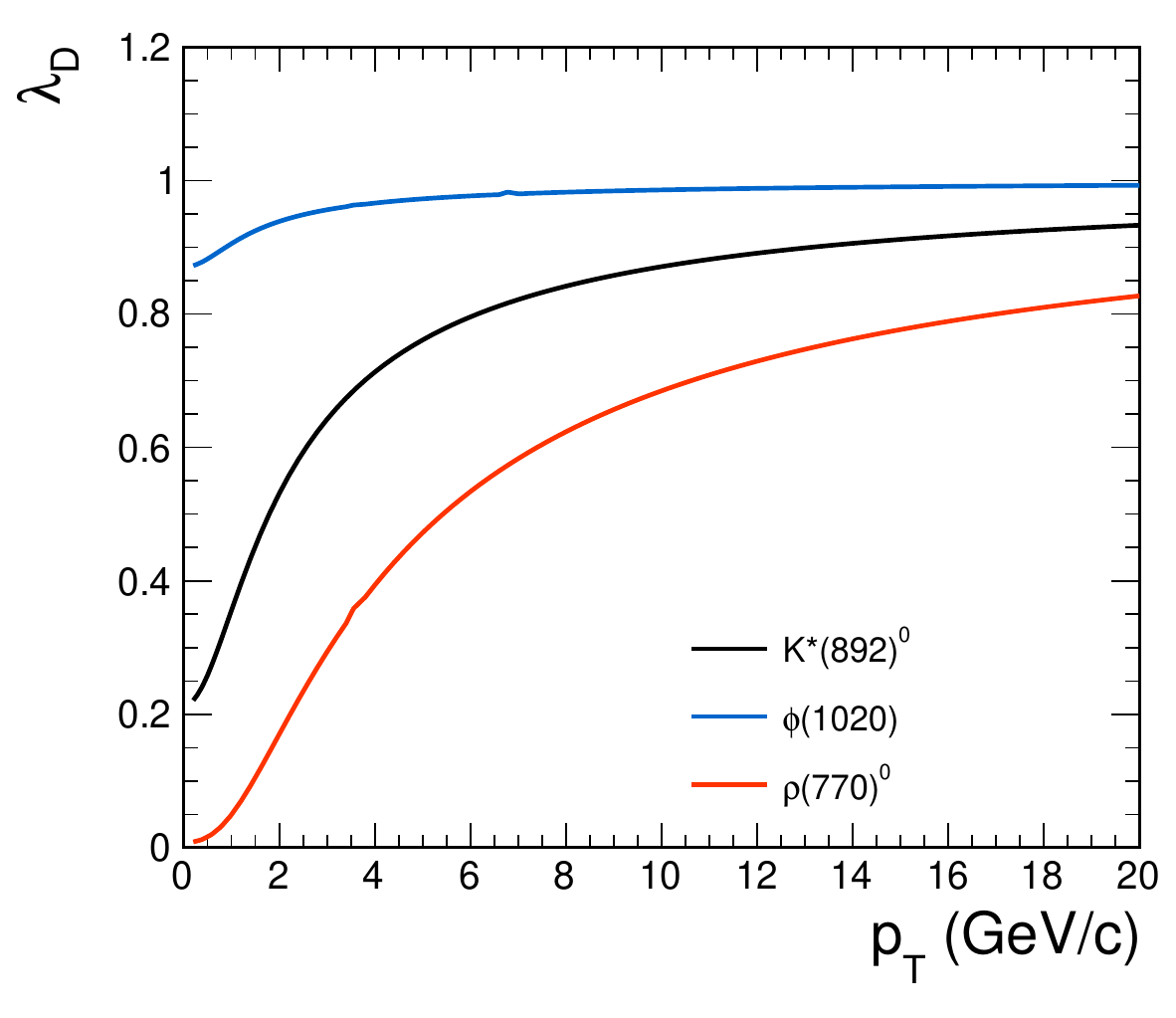}
\includegraphics[scale = 0.28]{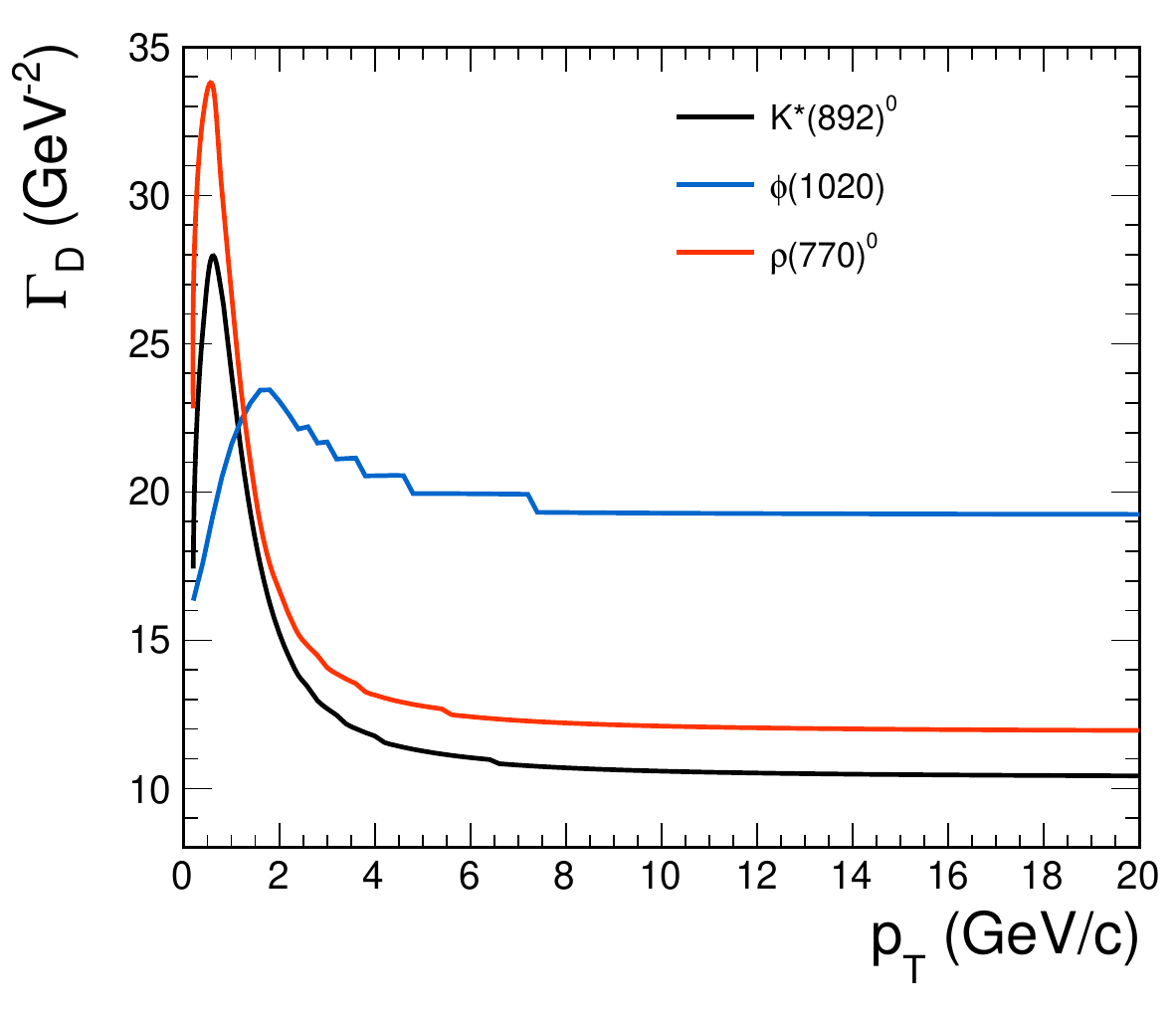}
\includegraphics[scale = 0.28]{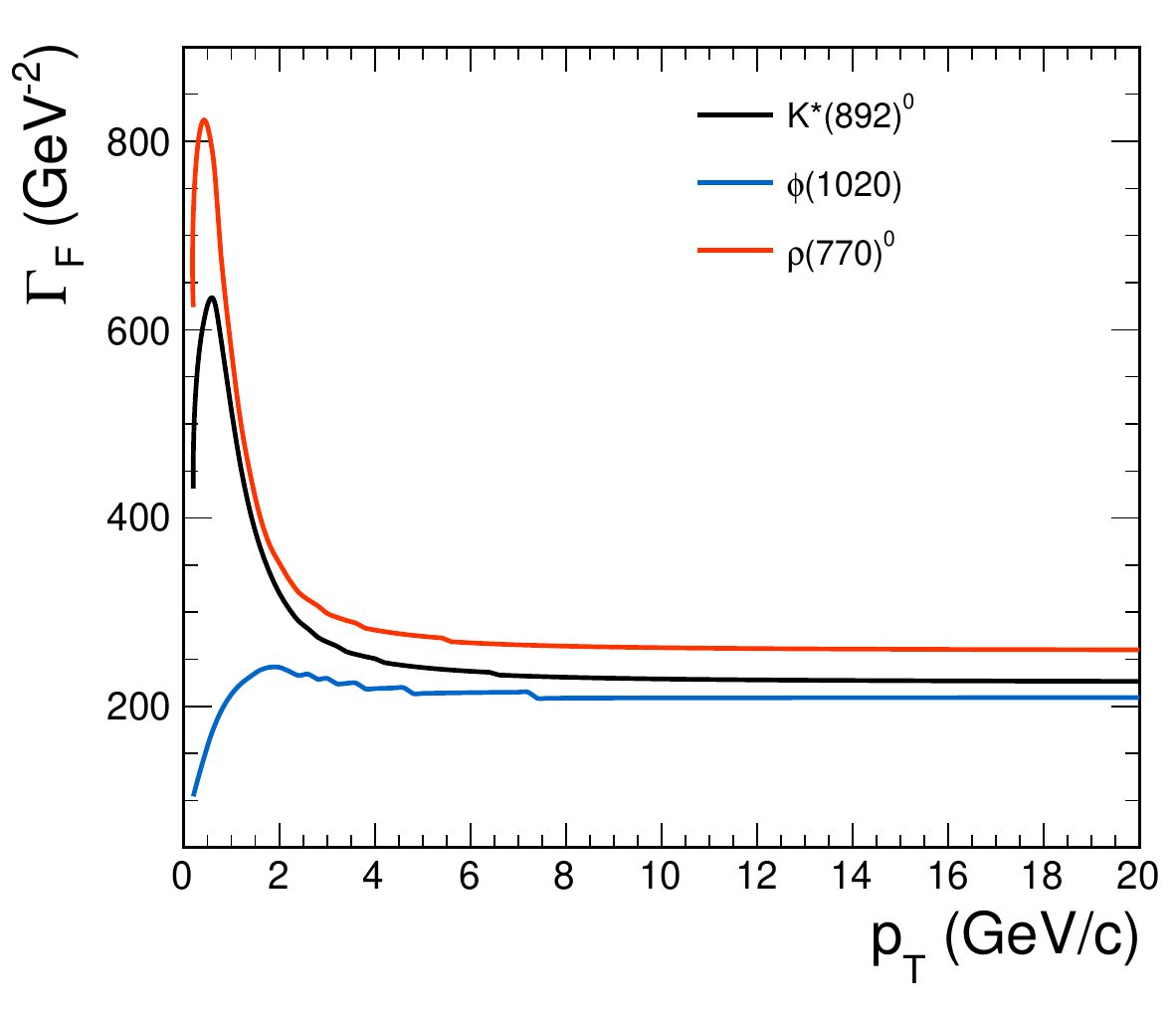}
\caption{The natural decay parameter $\lambda_{D}$ (\textbf{left}), dissociation reactivity $\Gamma_{D}$ (\textbf{middle}), and regeneration reactivity $\Gamma_{F}$ (\textbf{right}) explored as a function of transverse momentum ($p_{T}$) for $K^*(892)^0$, $\phi(1020)$, and $\rho(775)^0$.}
\label{fig3}
\end{figure*}

Further, we investigate the $p_T$-dependent natural decay, co-moving hadron-induced decay, and regeneration 
of the resonance particles within the hadronic phase as given by Equation~\eqref{eq22},~\eqref{eq24}, and~\eqref{eq25}, respectively. 
The results for the (0--5)\% centrality (most central) Pb--Pb collisions at the nucleon-nucleon center-of-mass energy 
$\sqrt{s_{NN}} = $ 5.02 TeV with these parameters are presented in Figure~\ref{fig3}.  
In Figure~\ref{fig3} (left), the natural decay rate is analyzed as a function of momentum for different resonance particles. At 
high $p_T$, the time dilation causes $\lambda_D$ to increase and approach unity. The impact of particle lifetime on the survival probability is evident, as 
the $\phi(1020)$ meson with $\tau_{reso} = $ 46.3 fm/c is relatively unaffected beyond $p_T = $ 2 GeV/c. In contrast, the yields of 
$\rho(770)^0$ mesons, which have $\tau_{reso} = $ 1.3 fm/c, are reduced over an extended $p_T$ range. The middle and right panels of 
Figure~\ref{fig3} displays the $p_T$-dependent interaction rates for co-mover-induced decay and regeneration, respectively. It is 
observed that decay and regeneration are more likely at low $p_T$ for all the considered resonances. These probabilities decrease 
further and saturate towards high $p_T$, which can be attributed to the initial particle production at these $p_T$ intervals. It is 
more probable for particles at low $p_T$ to move together, favoring the interaction, compared to particles with high $p_T$. In the 
following discussions, ``ND'', ``CMD'', and ``R'' mark natural decay, co-moving hadron-induced decay, and regeneration effects, respectively. The individual effects of the regeneration,  
decay due to co-moving hadrons, and natural decay are explicitly studied.

\begin{figure*}[htp]
\includegraphics[scale = 0.28]{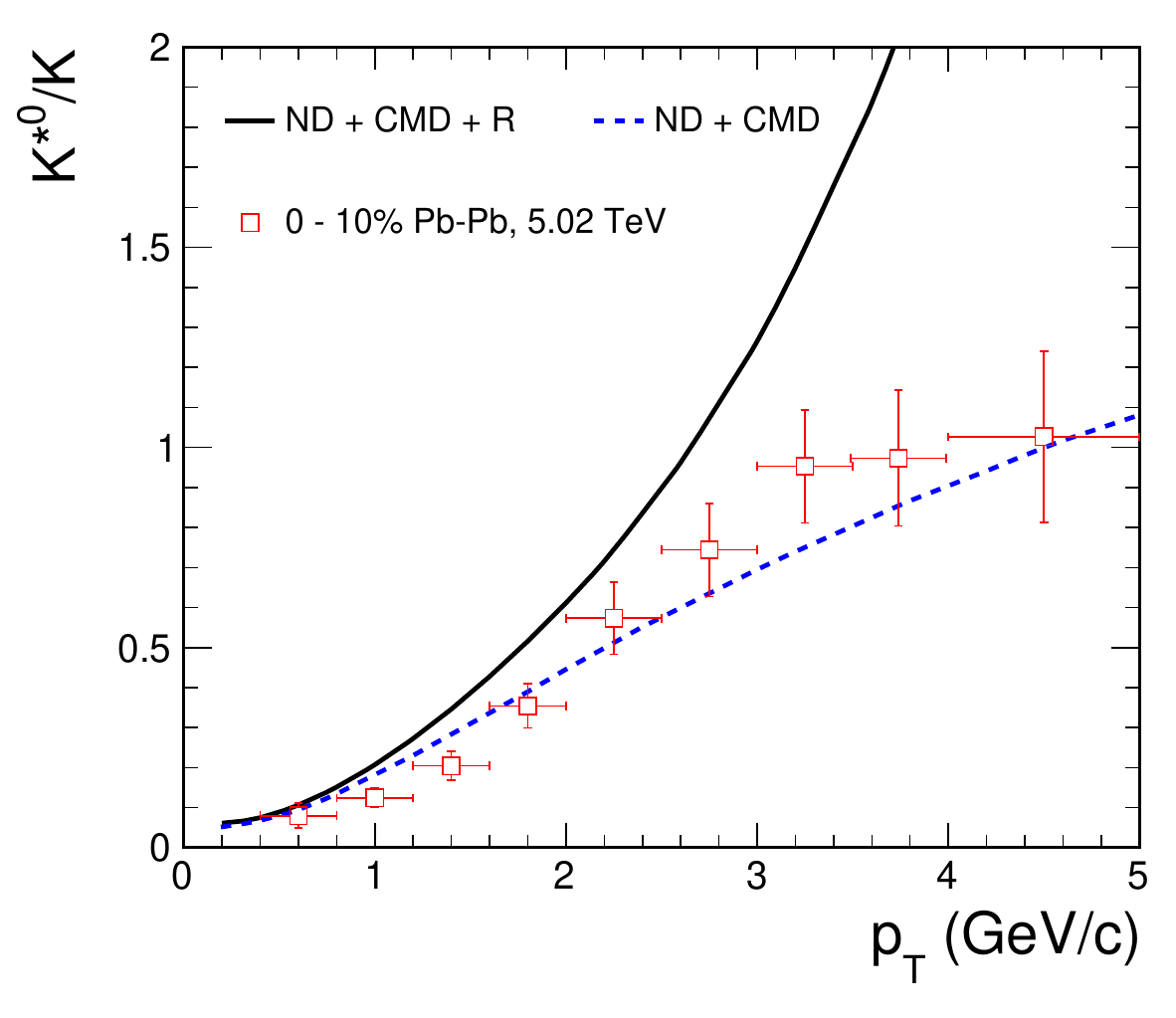}
\includegraphics[scale = 0.28]{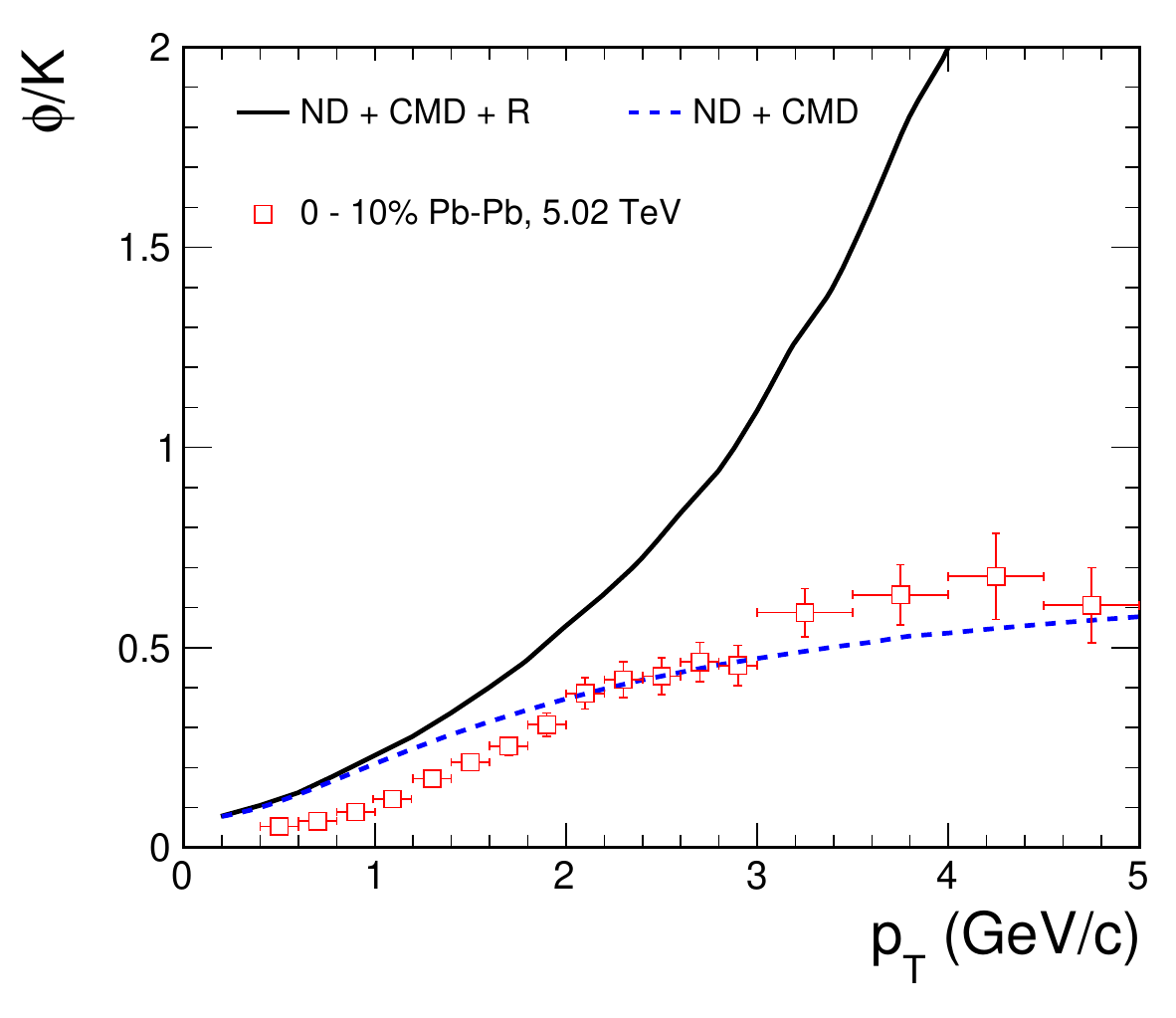}
\includegraphics[scale = 0.28]{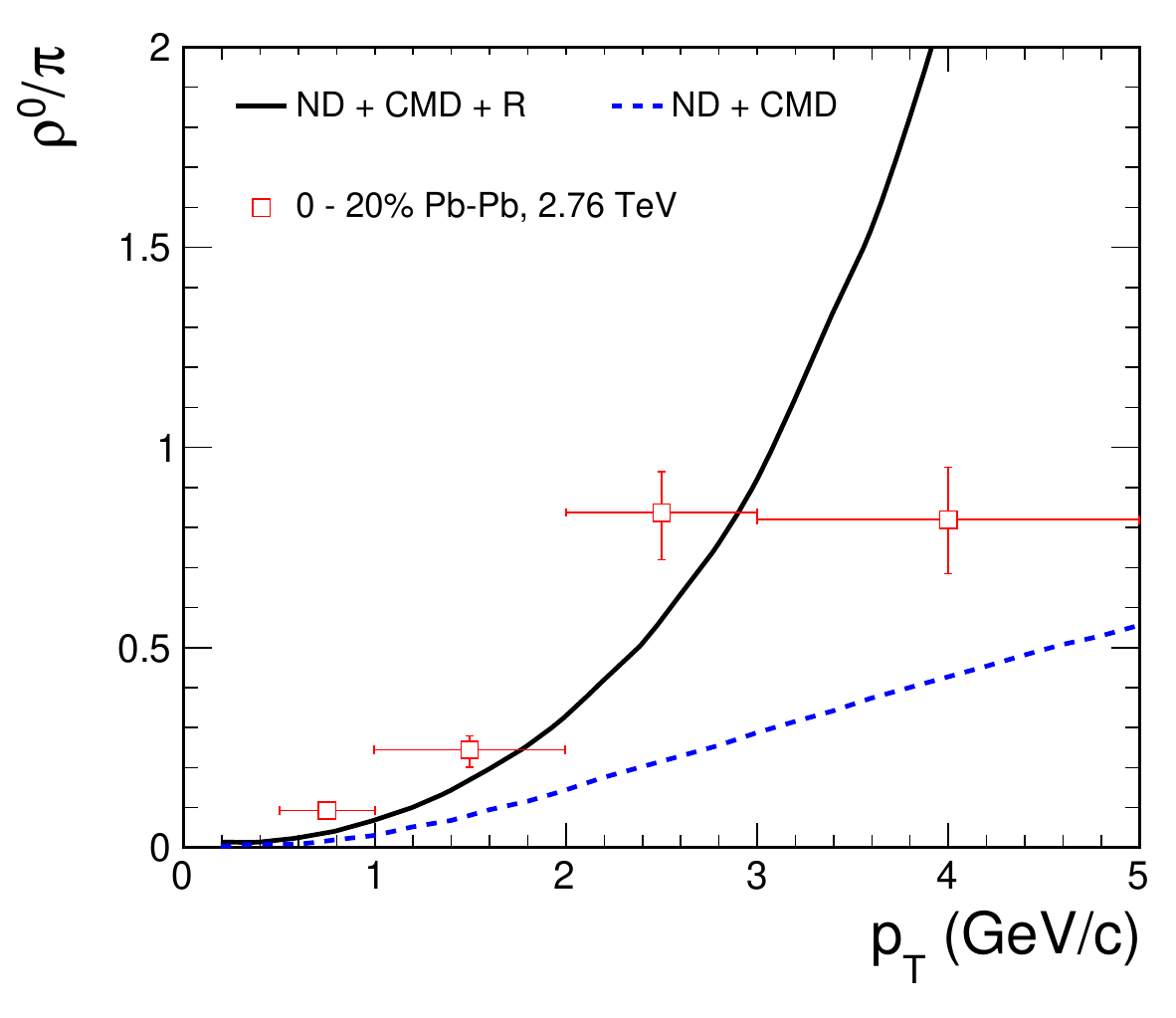}
\caption{Transverse-momentum ($p_{T}$) dependence of $K^*(892)^0/K$ (\textbf{left}), $\phi(1020)/K$ (\textbf{middle
}),  and $\rho(775)^0/\pi$ (\textbf{right}) ratios compared to the results from the ALICE experiment~\cite{ALICE:2019xyr, ALICE:2018qdv}. See text for details.}
\label{fig5}
\end{figure*}

The combined effects of these mechanisms on the resonance yields are further studied and shown in Figure~\ref{fig5} as a function of 
$p_T$. The particle ratios are shown only for the ``ND+CMD+R'' and ``ND+CMD'' cases, as the ``ND+R'' case is nearly identical to the 
``ND+CMD+R'' scenario. In all the three particle ratios considered, a qualitative agreement is obtained for $p_T < 3$ GeV for the 
``ND+CMD+R'' case beyond which the particle ratios are overestimated. However, a complete description of the full $p_T$ range is not 
expected due to the limitations of hydrodynamic calculations and its ability to account for high $p_T$ particles. In the absence of 
regeneration effects, these ratios tend to be much lower towards higher $p_T$. This indicates that our model overestimates 
regeneration effects towards the high $p_T$ region. Such an overestimation is mainly due to the effect of the initially obtained 
$p_T$-dependent daughter particle numbers, which was seen to overshadow the initial resonance particle production. Thus, the final 
particle numbers obtained at high $p_T$ tend to increase rapidly. A more complete description would be needed to account for these 
effects and obtain the saturating out of particle ratios to high $p_T$ seen experimentally.

\begin{figure*}[htp]
\includegraphics[scale = 0.28]{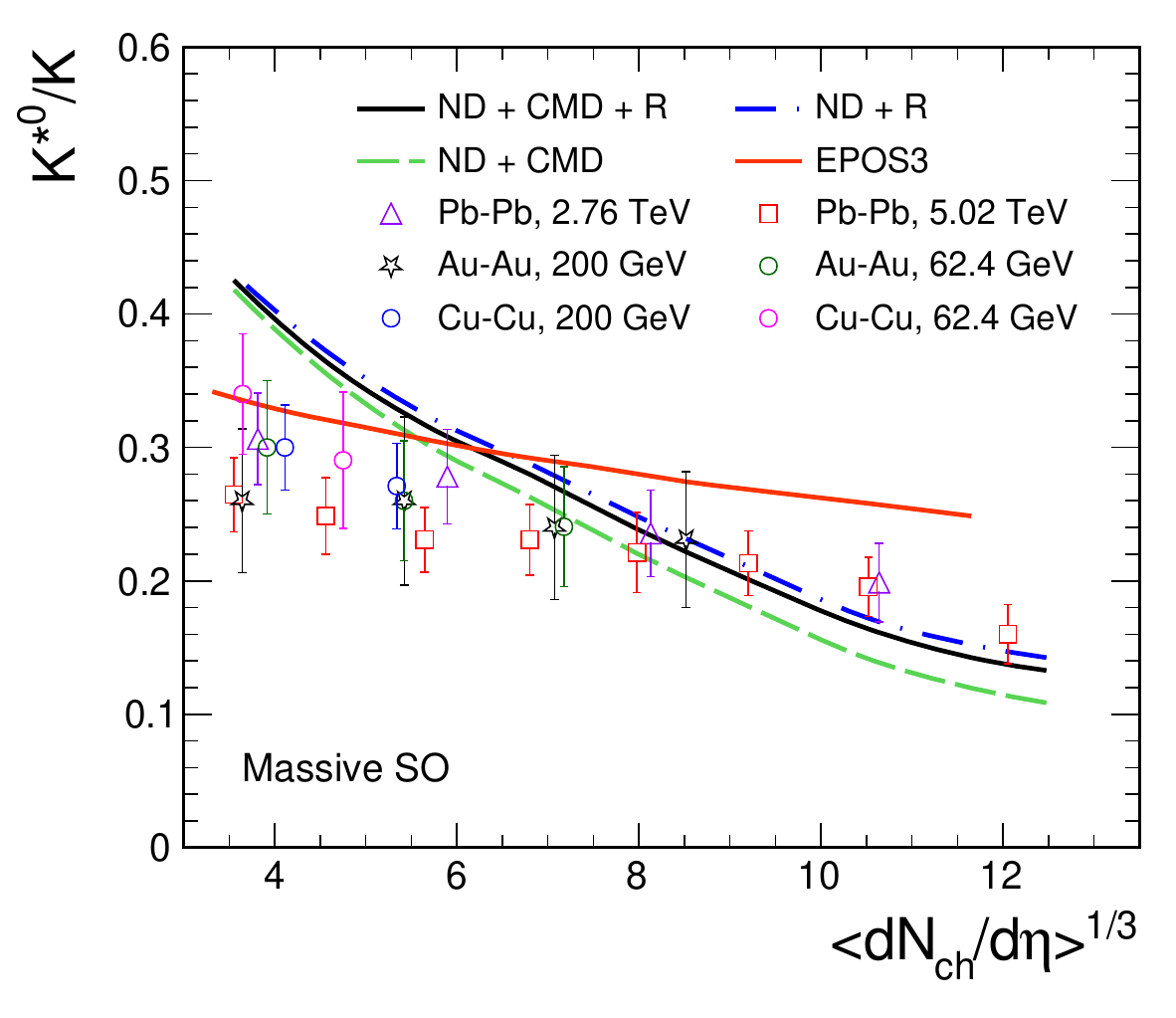}
\includegraphics[scale = 0.28]{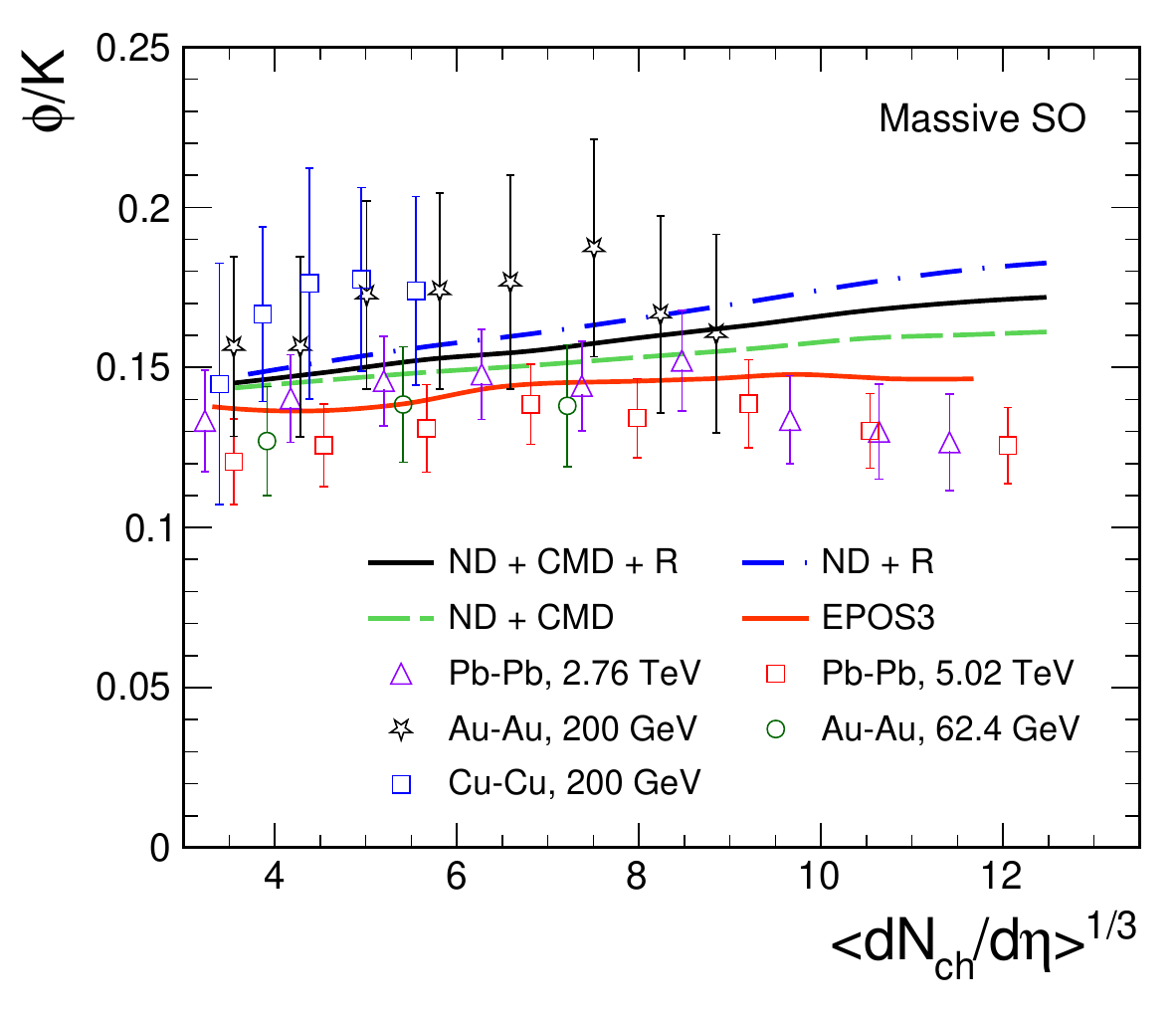}
\includegraphics[scale = 0.28]{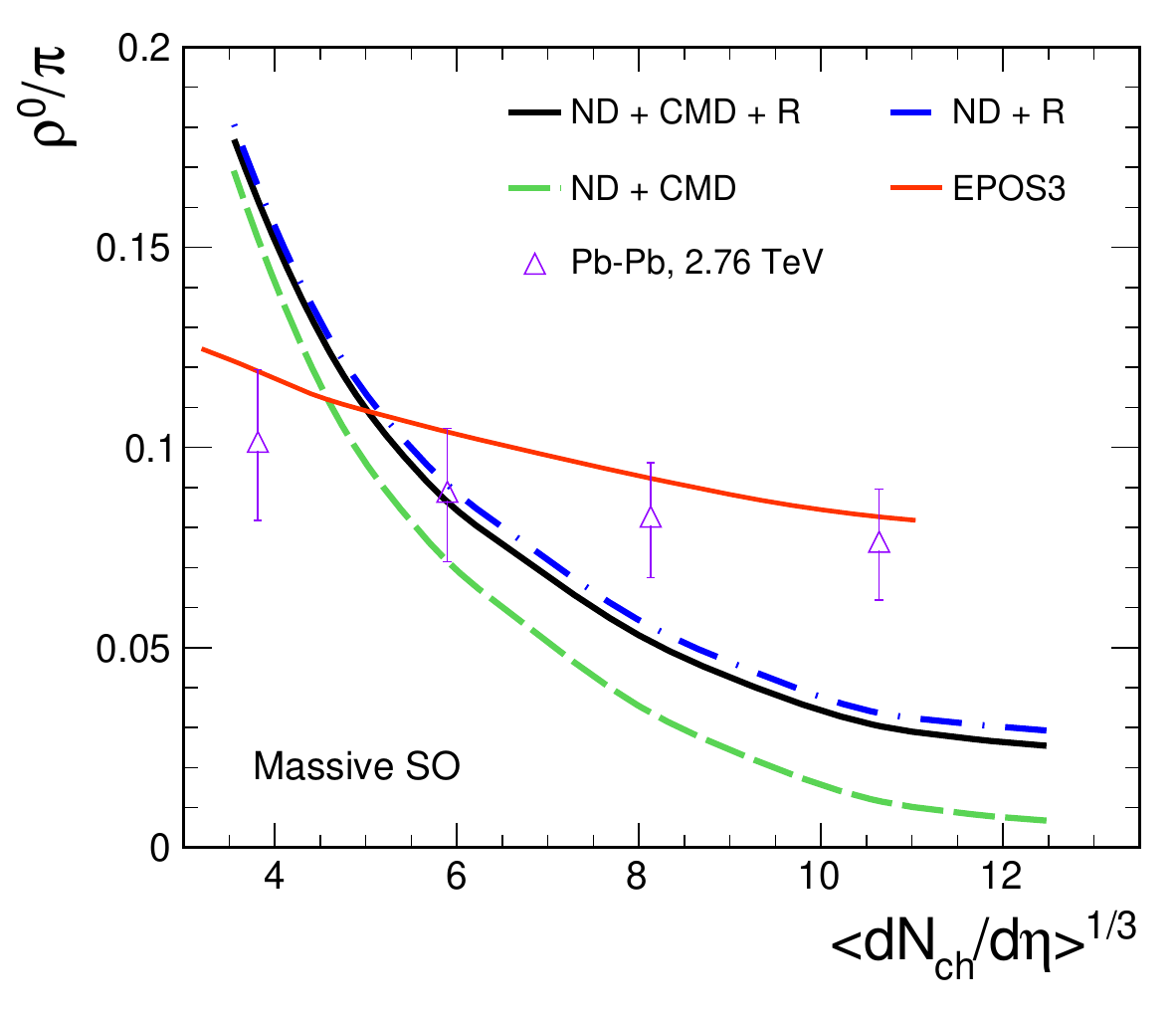}
\caption{The estimated ratios of $K^*(892)^0/K$ (\textbf{left}), $\phi(1020)/K$ (\textbf{middle
}), and $\rho(775)^0/\pi$ (\textbf{right}) obtained using 
 the calculations here compared to the data \cite{ALICE:2019xyr,ALICE:2014jbq, STAR:2010avo,STAR:2008bgi,ALICE:2018qdv} by the ALICE and RHIC experiments obtained with different colliding species. The results~\cite{ALICE:2019xyr,ALICE:2018qdv} from the EPOS3 model with UrQMD hadronic evolution are also shown for comparison. See text for details.}
\label{fig4}
\end{figure*}

Figure~\ref{fig4} depicts the final state particle ratios of resonances to stable particles obtained in heavy-ion collisions.  
Figure~\ref{fig4} compares the values obtained in our calculations for massive SO theory to the experimentally obtained $K^*(892)^0/K$, $\phi(1020)/K$ and $\rho(775)^0/\pi$ ratios ~\cite{ALICE:2019xyr, ALICE:2014jbq, STAR:2010avo, STAR:2008bgi, ALICE:2018qdv}. The massless SO case is not shown here as it does not account for a realistic scenario. Also, there is a small change in hadronic phase lifetime as compared to the massive case shown in Figure~\ref{fig2}. 
A considerable agreement with experimental values is obtained for $K^*(892)^0/K$ and $\phi(1020)/K$ estimates, even though 
$\sigma_{reg}^\phi(T)$ is approximated to be constant. At the same time, $\rho(775)^0/\pi$ values are underestimated, especially at the high multiplicity regions. The natural decay component plays a significant 
part in determining the trends obtained for $K^*(892)^0/K$ and $\rho(775)^0/\pi$ estimates due to their sufficiently small lifetimes. For $\phi(1020)/K$, with $\lambda_D$ being large, as seen in Figure~\ref{fig3}, there is quite a little change due to natural decay.
It can also be seen that both co-moving hadron-induced decay and regeneration effects play 
 significant roles at high multiplicity bins, while their impact on low multiplicities is marginal. 
Our study suggests that these effects must be considered while using resonances as a probe to study the hadronic phase produced in relativistic nuclear collisions.

\section{Summary}
\label{sum}
In this study, a hydrodynamic description is employed in the hadronic phase motivated through the dimensionless parameter, 
the Knudsen number, which implies the applicability of hydrodynamics. 
We consider a case where the regime between chemical freezeout and kinetic freezeout is dominated by hadronic scatterings but still maintains the applicability of hydrodynamics. We have found that the lifetime obtained through a hydrodynamic description of the hadronic phase is comparable with the results obtained at the ALICE experiment at LHC. Further, it is observed that the hadronic phase lifetime increases with the system size, which 
is manifested here in terms of charged-particle pseudorapidity density $\langle dN_{\rm ch}/d\eta\rangle$. We have shown that in heavy-ion collisions, the hadronic phase lifetime varies between 0.5 fm to 
 about 6 fm from ultra-peripheral to most-central collisions, respectively. We have also explored the yield modification of the resonance particles due to the natural decay and dissociation due to the co-moving hadrons during the hydrodynamic evolution of the hadronic phase. The regeneration of the resonances due to the combination of the required hadrons is also studied, and it is found that regeneration plays a crucial role besides dissociation mechanisms in obtaining the net yield of the resonance particles moving in the expanding system. 
The interplay between the decay and regeneration processes determines the final state yields of these particles.
The obtained results are compared with corresponding experimental data. These results qualitatively explain the experimental results at (low) $p_{T} \le$ 3 GeV as expected, because at high $p_{T}$, the applicability of the hydrodynamics is questionable. Further, $p_{T}$-integrated resonance particle ratios are obtained against charged particle multiplicity and compared with corresponding experimental results, showing considerable agreement.

Particle interactions in the hadronic phase significantly affect the final state particle yields~\cite{Andronic:2017pug}. These interactions can also affect particle correlations, leading to modifications to the corresponding values at the chemical freeze-out boundary~\cite{Pradhan:2021zbt}. Estimating the hadronic phase lifetime and decoupling the effects of these interactions is thus necessary to obtain precise information about the QCD fireball.
The present study explored the applicability of hydrodynamics, considering only massive pions as a component of the thermalized fluid using 0+1D viscous Bjorken-like flow. However, other abundant hadrons must be included as a part of the fluid, along with an extended 3+1D viscous hydrodynamics, for a comprehensive study. Such a comprehensive study can be performed with proper treatment of partial chemical equilibrium along the evolution. Furthermore, the estimates obtained here can have uncertainties due to the interaction cross-sections chosen from EVHRG, as the interaction cross-sections are poorly known near 
the chemical freeze-out boundary. These may be addressed by a proper estimation of the hadronic interaction cross-sections. 

\section*{Acknowledgement}
Ronald Scaria acknowledges CSIR, the Government of India, for the research fellowship. Raghunath Sahoo and Captain R. Singh  
acknowledge the financial support under DAE-BRNS, the Government of India, Project No. 58/14/29/2019-BRNS. The 
authors acknowledge the Tier-3 computing facility in the experimental high-energy physics laboratory of IIT Indore, supported by the ALICE project.



 \end{document}